\documentclass[journal]{IEEEtran} 

\usepackage{amsmath,amssymb,amsfonts,amsthm,color,fancybox,wrapfig,url,paralist,float,verbatim,varioref,overpic,wrapfig,multirow}
\usepackage{graphicx}

\usepackage[ruled,vlined]{algorithm2e}
\usepackage{caption}
\usepackage{subcaption}

\usepackage{tikz}
\usepackage[european resistor, european voltage, european current]{circuitikz}
\usetikzlibrary{arrows,shapes,positioning,snakes,calc}
\usetikzlibrary{decorations.markings,decorations.pathmorphing,decorations.pathreplacing}
\usetikzlibrary{calc,patterns,shapes.geometric,topaths,fit}

\makeatletter
\def\nl#1#2{\begingroup
    #2%
    \def\@currentlabel{#2}%
    \phantomsection\label{#1}\endgroup
}
\makeatother

{\ensuremath{
\begin{empheq}[box=\fbox]{align}
{#1}
\end{empheq}
}}

\newtheorem{theorem}            {Theorem}[section]

\newtheorem{definition}         [theorem]{Definition}

\newcommand{\normal}{\mathbf{N}}  

{
\begin{mdframed}
\par\noindent\textbf{#1:}\begin{rmfamily}\noindent}%
{\end{rmfamily}
\end{mdframed}
}

\newcommand{\hobsm}{\obsm}
\newcommand{\post}{\rho}

\newcommand{\aprob}{G}

\newcommand{\unpost}{q}
\newcommand{\unpostm}{\unpost^\model}
\newcommand{\horizon}{N}

\newcommand{\kg}{\psi}

\newcommand{\obsmt}{C^o}

\newcommand{\anoise}{\epsilon}

\newcommand{\pdf}{p}

\newcommand{\obs}{y}

\newcommand{\snoise}{w}
\newcommand{\onoise}{v}
\newcommand{\statem}{A}

\newcommand{\obsm}{C}

\newcommand{\snoisecov}{Q}
\newcommand{\onoisecov}{R}

\newcommand{\state}{x}
\newcommand{\statespace}{\mathcal{X}}
\newcommand{\obspace}{\mathcal{Y}}
\newcommand{\statedim}{X}
\newcommand{\obsdim}{{Y}}

\newcommand{\fun}{\phi}

\newcommand{\oprob}{B}
\newcommand{\tp}{P}

\newcommand{\finaltime}{N}

\newcommand{\model}{\theta}

\newcommand{\trace}{\operatorname{Tr}}

\newcommand{\Model}{\Theta}
\newcommand{\lik}{L_\finaltime} 

\newcommand{\probe}{\alpha}

\newcommand{\response}{\beta}
\newcommand{\dtime}{n}

\newcommand{\mle}{\model^*}

\newcommand{\belief}{\pi}

\newcommand{\Belief}{\Pi}

\newcommand{\modeldim}{M}

\newcommand{\enemystate}{\hat{\hat{\state}}}
\newcommand{\enemystatem}{\bar{\statem}}
\newcommand{\enemyobsm}{\bar{\obsm}}
\newcommand{\enemykalmancov}{\bar{\kalmancov}}

\newcommand{\enemySig}{\bar{\Sig}}
\newcommand{\enemyonoisecov}{\bar{\onoisecov}}
\newcommand{\enemysnoisecov}{\bar{\snoisecov}}
\newcommand{\enemyinputm}{\bar{F}}
\newcommand{\enemykg}{\bar{\kg}}

\newcommand{\eaction}{\bar{a}}

\makeatletter
\newcommand{\pushright}[1]{\ifmeasuring@#1\else\omit\hfill$\displaystyle#1$\fi\ignorespaces}
\makeatother

  {\popQED\end{theorem}}

\newcommand{\anoisecov}{\sigma^2_\anoise}

   \newcommand{\laction}{u}

\newcommand{\kalmancov}{\Sigma}

\newcommand{\Sig}{S}   

\newcommand{\imp }{\pi}

\newcommand{\filter}{T}

\newcommand{\argmax}{\operatornamewithlimits{argmax}}

\newcommand{\reals}{{\rm I\hspace{-.07cm}R}}

\newcommand{\beq}{\begin{equation}}
\newcommand{\eeq}{\end{equation}}

\newcommand{\p}{\prime}

\newcommand{\action}{a}

\def\param{{\theta}}

\newcommand{\innovations}{\iota}

\newcommand {\vect} {\textit{v}}

\newcommand{\hstate}{\hat{\state}}

\newcommand{\barray}{\begin{array}{ll}}
\newcommand{\earray}{\end{array}}

\newcommand{\lagrange}{\lambda}
\newcommand{\probedim}{m}

\newcommand{\sindx}{s}
\newcommand{\tindx}{t}

\newcommand{\utility}{U}

\newcommand{\dataset}{\mathcal{D}}
    \newcommand{\norm}[1]{\lVert#1\rVert}

\newcommand{\precconstraint}{p_*}
\renewcommand{\vect}[1]{\pmb{#1}}

\newcommand{\sense}{\eta}

\usepackage{quoting}
\quotingsetup{vskip=1pt}

\tikzset{
    block/.style={rectangle, draw, line width=0.5mm, black, text width=4.5em, text centered,
                 minimum height=1em},
               line/.style={draw, -latex}}

      \tikzset{
    block3/.style={rectangle, draw, line width=0.5mm, black, text width=7.5em, text centered,
                 minimum height=1em},
               line/.style={draw, -latex}}

\tikzset{
    block2/.style={rectangle, draw, line width=0.5mm, text centered,
                 minimum height=2em},
               line/.style={draw, -latex}}

\tikzset{
    blocka/.style={rectangle, draw, line width=0.5mm, black, text width=4.5em, text centered,
                 minimum height=1em},
               line/.style={draw, -latex}}
               
\tikzset{
    blockb/.style={rectangle, draw, line width=0.5mm, black, text width=6em, text centered,
                 minimum height=1em},
               line/.style={draw, -latex}}

             \title{Adversarial Radar Inference:  Inverse Tracking, Identifying Cognition and Designing
              Smart Interference\thanks{This research was supported in part by the US Army Research Office under  grants
 W911NF-21-1-0093 and W911NF-19-1-0365, and the National Science Foundation
under grant CCF-1714180.}}
\author{Vikram Krishnamurthy,~\IEEEmembership{Fellow,~IEEE}, Kunal Pattanayak,~\IEEEmembership{Student Member,~IEEE}, Sandeep Gogineni,~\IEEEmembership{Member,~IEEE}, Bosung Kang,~\IEEEmembership{Member,~IEEE}, Muralidhar Rangaswamy,~\IEEEmembership{Fellow,~IEEE}
	\thanks{V. Krishnamurthy and K. Pattanayak are with the School
		of Electrical and Computer Engineering, Cornell University, Ithaca,
		NY, 14853 USA. e-mail: vikramk@cornell.edu, kp87@cornell.edu.
		
		S. Gogineni is with the Information Systems Laboratories, Inc., San Diego, CA, USA. e-mail: sgogineni@islinc.com.
		
		B. Kang is with the University of Dayton Research Institute, Dayton, OH, 45469 USA. e-mail: Bosung.Kang@udri.udayton.edu.
		
		M. Rangaswamy is with the Air Force Research Laboratory, Wright Patterson Air Force Base, OH, 45433 USA. e-mail: Muralidhar.Rangaswamy@us.af.mil.
		}}
\begin{document}

\maketitle
\begin{abstract} 
This paper considers three inter-related adversarial inference problems involving cognitive radars. We first discuss inverse tracking of the radar to estimate the adversary's estimate of us based on the radar's actions and calibrate the radar's sensing accuracy. Second, using revealed preference from microeconomics, we formulate a non-parametric test to identify if the cognitive radar is a constrained utility maximizer with signal processing constraints. We consider two radar functionalities, namely, beam allocation and waveform design, with respect to which the cognitive radar is assumed to maximize its utility and construct a set-valued estimator for the radar's utility function. Finally, we discuss how to engineer interference at the physical layer level to confuse the radar which forces it to change its transmit waveform.
The levels of abstraction range from smart interference design based on Wiener filters (at the pulse/waveform level), inverse Kalman filters at the tracking level and revealed preferences for identifying utility maximization at the systems level.
\end{abstract}

\subsection*{Glossary of Symbols.}  
\begin{tabular}{cl}
\multicolumn{2}{l}{{\bf Inverse Tracking (Sec.\,\ref{sec:cspomdp}) }}\\
$\state_k$ & our kinematic state at time $k$ \\
$\tp_{\state_{k+1},\state_k}$ & transition kernel $p(\state_{k+1}|\state_k)$\\
$\snoise_k$ & state noise at time $k$\\
$\snoisecov_k$ & covariance of $\snoise_k$\\
$\obs_k$ & observation of $\state_k$ \\
$\onoise_k$ & observation noise at time $k$ \\ 
$\onoisecov_k$ & covariance of $\onoise_k$\\
$\obsm$ & adversary sensor gain\\
$\oprob_{\state_k,\obs_k}$ & conditional pdf of $\obs_k$ given state $\state_k$ $p(\obs_k|\state_k)$ \\
$\belief_k$ & adversary's belief of $\state_k$\\ 
$T(\belief,\obs)$ & belief update\\
$\hstate_k$ & conditional mean of state estimate of $\state_k$\\
$\kalmancov_k$ & covariance of state estimate of $\state_k$\\
$\laction_k$ & adversary's action at time $k$ \\ 
$\fun$ & stochastic mapping from $\belief_k$ to $\laction_k$ \\
$\action_k$ & our measurement of $\laction_k$ \\ 
$\aprob_{\belief_k,\action_k}$ & conditional pdf of $\action_k$ given belief $\belief_k$  $p(\action_k|\belief_k)$\\
$\post_k$ & our belief of $\pi_k$ given $\state_k$ and $\action_k$ \\
$\enemystate_k$ & conditional mean of $\hstate_k$\\
$\enemykalmancov_k$ & covariance of $\hstate_k$ \\
$\param$ & model parameter for $\obsm$ \\
$\lik$ & log-likelihood \\ 
$\innovations_k$ & innovations of inverse Kalman filter\\
\multicolumn{2}{l}{{\bf Identifying Cognition (Sec.\,\ref{sec:RP_UM}) }}\\
$n$ & slow time scale index \\
$\probe_n$ & probe signal at time $n$ \\
$\response_n$ & adversary's response at time $n$ \\
$\utility(\response)$ & adversary's utility function\\
\multicolumn{2}{l}{{\bf Smart Interference (Sec.\,\ref{sec:Interference}) }}\\
$\vect X_l$ & adversary radar's received signal at $l^{\text{th}}$ pulse\\
$\vect H_t(l)$ & transmit channel impulse response at $l^{\text{th}}$ pulse\\
$\vect H_c(l)$ & clutter channel impulse response at $l^{\text{th}}$ pulse\\
$\vect H_p(l)$ & probe signal at $l^{\text{th}}$ pulse\\
$\vect W(l)$ & radar's transmission waveform at $l^{\text{th}}$ pulse\\
$\vect E_r(l)$ & measurement noise in $\vect X(l)$ \\ 
$C_r$ & covariance of $\vect E_r(l)$\\
$\vect Y(l)$ &  our observation of $\vect W(l)$\\
$\vect E_o(l)$ &  measurement noise in $\vect Y(l)$\\
$C_o$ & covariance of $\vect E_o(l)$\\
\end{tabular}

\section{Introduction}
Cognitive sensors are  reconfigurable sensors that optimize  their sensing mechanism and transmit functionalities. The concept of cognitive radar~\cite{CKH09,KD07,KD09,Hay12} has evolved over the last two decades  and a common aspect is the sense-learn-adapt paradigm. A cognitive fully adaptive radar  enables joint optimization of the adaptive transmit and receive functions by sensing (estimating) the radar channel that includes clutter and other interfering signals \cite{BGG15,GBG16}.

\subsection{Objectives} \label{sec:objectives}
\begin{figure*}
    \centering
            {\resizebox{13cm}{!}{
\begin{tikzpicture}[node distance = 1cm, auto]
    \node [blockb] (BLOCK1) {Sensor \\ (Receiver)};
    \node [blockb, below of=BLOCK1,right of=BLOCK1,node distance=1.5cm] (BLOCK2) {Decision Maker};
    \node [blockb, below of=BLOCK1,left of=BLOCK1,node distance=1.5cm] (BLOCK3) {Tracker \\ (Estimator)};

    \draw[<-] (BLOCK1) -|   (BLOCK2)  ;
    \draw[->] (BLOCK1.west) -|    (BLOCK3);

    \draw[->](BLOCK3) --  (BLOCK2);
    
    \node [blockb,right of=BLOCK1,node distance = 8.5cm] (BLOCK4) {Inverse Tracking \\ (Sense)};
    \node [blockb, below of=BLOCK4,right of=BLOCK4,node distance=1.5cm] (BLOCK5) {Engineered Interference\\(Adapt)};
    \node [blockb, below of=BLOCK4,left of=BLOCK4,node distance=1.5cm] (BLOCK6) {Identifying Cognition \\ (Learn)};

    \draw[<-] (BLOCK4) -|  (BLOCK5)  ;
    \draw[->] (BLOCK4.west) -|  (BLOCK6);

    \draw[->] (BLOCK6) --  (BLOCK5);

    \draw[->,line width=2pt] ([xshift=0.4cm]BLOCK2.east)   -- ([xshift=-0.4cm]BLOCK6.west);
    \draw[->,line width=2pt] ([xshift=-0.4cm,yshift=1cm]BLOCK6.west) -- ([xshift=0.4cm,yshift=1cm]BLOCK2.east);
    \node[draw] at (8.5,-3) {Us (Counter-Adversarial System)};
    \node[draw] at (0,-3) {Cognitive Radar (Adversary)};
   \end{tikzpicture}} }

\caption{Schematic illustrating the main ideas in the paper. The three components on the right are inter-related and constitute the sense-learn-adapt paradigm of the observer~(``us'') reacting to a reactive system such as the cognitive radar (on the left). This paper considers the above schematic 
and proposes counter-adversarial schemes against cognitive radars 
for different levels of abstraction, i.e.\,, interference design based on Wiener filters at the pulse/waveform level, inverse Kalman filters at the Bayesian tracking level, and revealed preference techniques for estimating adversary's utility function at the systems level.}
\label{fig:blockdiag}
\end{figure*}
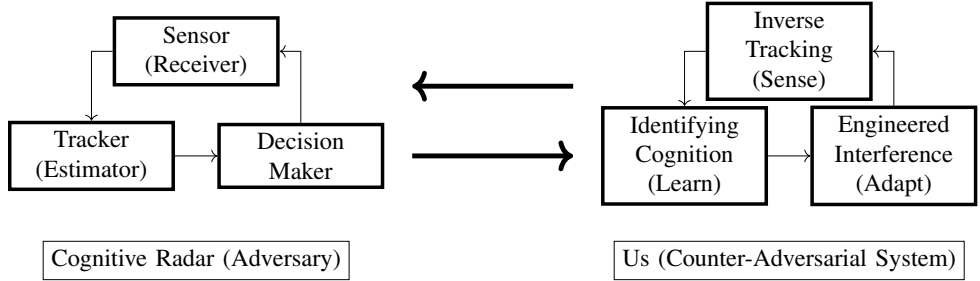

This paper addresses the next step and achieves the following objectives schematically shown in Figure \ref{fig:blockdiag}.
   The framework in this paper involves an adversarial signal processing problem  comprising   ``us'' and an ``adversary''.
``Us'' refers to an asset such as a drone/UAV  or electromagnetic signal that probes an ``adversary'' cognitive radar.
 Figure~\ref{fig:graph} shows the schematic setup. A cognitive sensor  observes our kinematic state $\state_k$ in noise as the observation $\obs_k$. It then  uses a Bayesian tracker to update its posterior distribution $\belief_k$ of our state $\state_k$ and chooses an action $\laction_k$ based on this posterior. We observe the sensor's action in noise as $\action_k$. 
Given knowledge of ``our'' state sequence  $\{\state_k\}$  and the observed    actions $\{\action_k\}$ taken by the adversary's sensor, 
we  focus on the following   inter-related aspects:

 {\bf  1. Inverse tracking and estimating the Adversary's Sensor Gain}.
  Suppose the adversary radar observes our state in noise; updates its posterior  distribution $\belief_k$ of our state $\state_k$ using a Bayesian tracker, and then chooses an action $\laction_k$ based on this posterior. Given knowledge of ``our'' state and sequence of noisy measurements $\{\action_k\}$ of the adversary's actions $\{\laction_k\}$, how can the adversary radar's posterior distribution (random measure) be estimated? We will develop an inverse Bayesian filter for tracking the radar's posterior belief of our state and present an example involving the Kalman filter where the inverse filtering problem admits a finite dimensional characterization.

A related question is:
How to remotely estimate  the adversary radar  sensor's conditional pdf of observation given the state when it  is estimating us?  This is important because it tells us how accurate the adversary's sensor is; in the context of Figure \ref{fig:graph} it tells us, how accurately the adversary tracks our drone. The data we have access to is our
 state (probe signal) sequence $\{\state_k\}$ and  measurements of the adversary's radar actions $\{\action_k\}$.
Estimating the adversary's sensor accuracy is  non-trivial with several challenges.
   First, even though we know our state and state dynamics model (transition law),
 the adversary does not.
 The adversary  needs to estimate our state and state transition law  based on our trajectory; and we  need to estimate the adversary's estimate of our
 state transition law. Second, computing the MLE of the adversary's sensor gain also requires inverse filtering.

  {\bf 2. Revealed Preferences and Identifying Cognitive Radars.} 
Suppose the cognitive radar is a constrained utility maximizer that optimizes its actions $\laction_k$ subject to physical level (Bayesian filter) constraints. How can we detect this utility maximization behavior? The actions $\laction_k$ can be viewed as resources the radar adaptively allocates to maximize its utility.
We consider two such resource allocation problems, namely,
  \begin{compactitem}
      \item {\em Beam Allocation}: The radar adaptively switches its beam while tracking multiple targets.  
      \item {\em Waveform Design}: The radar adaptively designs its waveform while ensuring the signal-to-interference-plus-noise ratio~($\operatorname{SINR}$) exceeds a pre-defined threshold. 
  \end{compactitem}
Nonparametric detection of utility maximization behavior is the central theme of {\em revealed preference} in microeconomics. A remarkable result is \textit{Afriat's theorem}: it  provides a necessary and sufficient condition for a finite dataset  to
 have originated from a utility maximizer. 
 We will  develop constrained set-valued utility estimation methods that  account for signal processing constraints introduced by the Bayesian tracker for performing adaptive beam allocation and waveform design respectively. 

{\bf 3. Smart Signal Dependent Interference.} 
We next consider the adversary radar choosing its transmit waveform for target tracking by implementing a Wiener filter to maximize its signal-to-clutter-plus-noise ratio~($\operatorname{SCNR}$\footnote{The terms $\operatorname{SCNR}$ and $\operatorname{SINR}$ are used interchangeably in the paper.}). By observing the optimal waveform chosen by the radar, our aim is to develop a  strategy to estimate the adversary cognitive radar channels and then construct  signal dependent interference generation  to confuse the adversary radar.

\subsection{Perspective}
The adversarial dynamics considered in this paper fit naturally within the so called Dynamic Data and Information Processing (DDIP) paradigm.
The adversary's radar  senses, adapts and learns from us. In turn we adapt, sense and learn from the adversary. So in simple terms we are modeling and analyzing the interaction of two DDIP systems. In this context, this paper has three major themes schematically shown in Figure \ref{fig:blockdiag}: inverse filtering which is a Bayesian framework for interacting DDIP systems, inverse cognitive sensing which is a non-parametric approach for utility estimation for interacting DDIP systems, and interference design to confuse the adversarial DDIP system.

This work is also  motivated by the design of counter-autonomous systems: given measurements of  the actions of an  autonomous adversary, how can our counter-autonomous system  estimate the underlying belief of the adversary, identify if the adversary is cognitive (constrained utility maximizer) and design appropriate probing signals to confuse the adversary. 
This paper generalizes and contextualizes recent works in adversarial signal processing \cite{KR19,KAEM20} which only deal with specific radar functionalities. Instead, this paper views the cognitive radar as a holistic system operating at three stages of sophistication 
unifies the three inter-related aspects of adversarial signal processing, namely, inverse tracking, identifying cognition and designing interference. The three components complement one another and constitute this paper's adversarial signal processing sense-learn-adapt (SLA) paradigm of Figure~\ref{fig:blockdiag}. 

\subsection{Organization} We conclude this section with a brief outline of the key results of the following sections, and their relevant to the sense, learn and adapt elements of the SLA paradigm of Figure~\ref{fig:blockdiag}.

{\bf Sense:} In Sec.\,\ref{sec:cspomdp}, we discuss inverse tracking techniques to estimate the sensor accuracy of an adversary radar. We focus mainly on the inverse Kalman filter and illustrate in carefully chosen examples how the adversary sensor's accuracy can be estimated. This constitutes the `sensing' aspect of the SLA paradigm.

{\bf Learn:} In Sec.\,\ref{sec:RP_UM}, we abstractly view the adversarial radar as a cognitive decision maker that maximizes a utility function subject to physical resource constraints. Specifically, we show that if the cognitive radar optimizes its waveform to maintain its SINR above a threshold, then we can identify (and hence, `learn') the utility function of the radar. The utility function provides deeper knowledge of the radar's behavior and constitutes the `learn' element of the SLA paradigm. 

{\bf Adapt:} In Sec.\,\ref{sec:Interference}, we consider a slightly modified setup where the radar chooses its waveform to maximize its SCNR. We show that by intelligently probing the radar with interference signals and observing the changes in the radar's waveform, we can confuse the adversary's radar by decreasing its $\operatorname{SCNR}$. This adaptive signal processing algorithm is justified only if the `sense' and `learn' aspect of the SLA paradigm function properly, that is, the counter-adversarial system knows how the radar will react to changes in its environment.

Finally, we emphasize that the three main aspects of inverse tracking (sensing the estimate of the adversary), identifying utility maximization (learning the adversary's utility function) and adaptive interference (adapting our response) are instances of the general paradigm of sense-learn-adapt in counter-adversarial systems.  As mentioned above, our formulation deals with the interaction of two such sense-learn-adapt systems.

\section{Inverse Tracking and Estimating Adversary's Sensor}  \label{sec:cspomdp}

This section discusses inverse tracking in an adversarial system schematically illustrated in Figure \ref{fig:graph}. Our main ideas involve estimating the adversary's estimate of us and estimating the adversary's sensor conditional pdf of observation given the state.

\subsection{Background and Preliminary Work}

    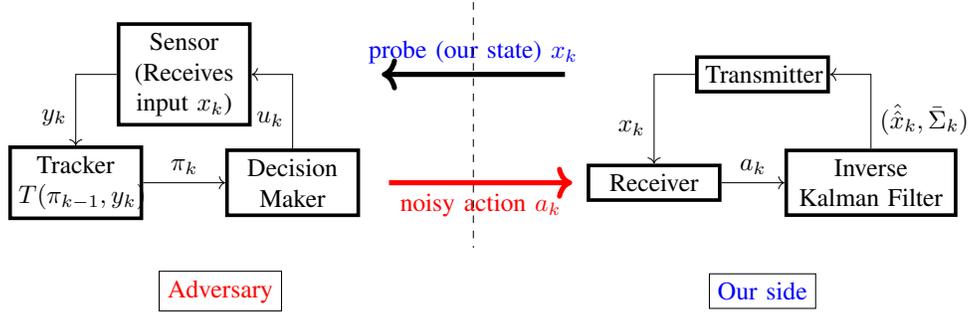
\begin{figure*}
    \centering
           {\resizebox{13cm}{!}{
\begin{tikzpicture}[node distance = 1cm, auto]
    \node [block] (BLOCK1) {Sensor\\
    (Receives input $\state_k$)};
    \node [block, below of=BLOCK1,right of=BLOCK1,node distance=1.5cm] (BLOCK2) {Decision \\ Maker};
    \node [block, below of=BLOCK1,left of=BLOCK1,node distance=1.5cm] (BLOCK3) {Tracker $\filter(\belief_{k-1},\obs_k)$};

    \draw[<-] (BLOCK1) -| node[left,pos=0.8]{$\laction_k$}  (BLOCK2)  ;
    \draw[->] (BLOCK1.west) -|   node[left,pos=0.8]{$\obs_k$} (BLOCK3);

    \draw[->](BLOCK3) --  node[above]{$\belief_k$} (BLOCK2);
    
     \node [block,right of=BLOCK1, node distance=8cm] (BLOCK4) {Transmitter};
    \node [blockb, below of=BLOCK4,right of=BLOCK4,node distance=1.5cm] (BLOCK5) {Inverse Kalman Filter};
    \node [block, below of=BLOCK4,left of=BLOCK4,node distance=1.5cm] (BLOCK6) {Receiver};

    \draw[<-] (BLOCK4) -| node[right,pos=0.8]{$(\hat{\hat{\state}}_k,\bar{\Sigma}_k)$}  (BLOCK5)  ;
    \draw[->] (BLOCK4.west) -|   node[left,pos=0.8]{$\state_k$} (BLOCK6);

    \draw[->](BLOCK6) --  node[above]{$\action_k$} (BLOCK5);


    \draw[->,color=black,line width=2pt] ([xshift=-1.8cm]BLOCK4.west)   -- node[above]{\color{blue}{probe (our state) $\state_k$}}([xshift=1.8cm]BLOCK1.east);
    \draw[->,color=red,line width=2pt] ([xshift=0.4cm]BLOCK2.east)   --   node[below]{noisy action $\action_k$} ([xshift=-0.2cm]BLOCK6.west);
    \node[draw] at (8,-3.0) {{\color{blue}Our side}};
    \node[draw] at (0.5,-3.0) {{\color{red}Adversary}};
    
\draw[dashed] (4,1) -- (4,-2.4);
  \end{tikzpicture}} 
  }

\caption{Schematic of Adversarial Inference Problem. Our side is a drone/UAV or electromagnetic signal that probes the  adversary's cognitive radar system. Based on the action $\action_k$ of the adversary, our side computes the estimate of the adversary's estimate of our state $\state_k$ using the inverse Kalman filter outlined in Sec.\,\ref{sec:inv_kalman}.}
\label{fig:graph}
\end{figure*}

We start by formulating the problem which involves two entities;  ``us'' and ``adversary''.  With $k=1,2,\ldots$ denoting discrete time, the model has
the following dynamics:
\begin{equation}
\begin{split}
    \state_k &\sim  \tp_{\state_{k-1},\state} = \pdf(\state | \state_{k-1}), \quad \state_0 \sim \belief_0 \\
    \obs_k  &\sim \oprob_{\state_k,\obs} = \pdf(y | x_k)\\
    \belief_k &= \filter(\belief_{k-1}, \obs_k)= \pdf(x_k| \obs_{1:k})\\
    \action_k &\sim \aprob_{\belief_k,\action} = \pdf(\action | \belief_k)
  \end{split} \label{eq:model}
\end{equation}
Let us explain the notation in (\ref{eq:model}):
\begin{compactitem}
    \item $\state_k\in \statespace$ is our Markovian state with  transition kernel $\tp_{\state_{k-1},\state}$,  prior $\belief_0$ and state space  $\statespace$.
    \item $\obs_k$ is the adversary's noisy observation of our state $\state_k$; with conditional pdf of observation given the state $\oprob_{\state_k,\obs}$.
    \item $\belief_k$ is the adversary's belief (posterior)  of our state $\state_k$ where $\obs_{1:k}$ denotes the observation sequence  $\obs_1,\ldots,\obs_k$. The operator $T$ in (\ref{eq:model}) is the classical Bayesian optimal  filter that computes the posterior belief of the state given observation $\obs$ and current belief $\belief$.
    \beq  \hspace{-0.8cm}\filter(\belief,\obs) = \operatorname{vec}\left( \frac{
    \oprob_{\state,\obs} \int_\statespace 
    \tp_{\zeta, \state}\, \belief(\zeta) \,d\zeta}
  {\int_\statespace  \oprob_{\state,\obs} \int_\statespace 
    \tp_{\zeta, \state}\, \belief(\zeta)\, d\zeta d\state},\state\in\statespace \right)
  \label{eq:belief}
\eeq
    
Let $\Belief$ denote the  space of all such beliefs. When
the state space
$\statespace$ is finite, then $\Belief$ is  the unit $
\statedim-1$ dimensional simplex of $\statedim$-dimensional probability mass functions.
    \item $\action_k $ denotes our measurement of the adversary's action based on its current belief $\belief_k$. The adversary chooses an action $\laction_k$ as a (possibly) stochastic function of $\belief_k$ and we obtain a noisy measurement of $\laction_k$ as $\action_k$. We encode this as $\aprob_{\belief_k,\action_k}$, the conditional probability of observing  action $\action_k$ given the adversary's belief $\belief_k$. Although not explicitly shown, $\aprob$ abstracts two stochastic maps: 1) the map from the adversary's belief $\belief_k$ to its action $\laction_k$, and 2) the map from the adversary's action $\laction_k$ to our noisy measurement $\action_k$ of this action.
    \end{compactitem}

Figure \ref{fig:graph} displays a schematic and graphical representation of the model (\ref{eq:model}). The schematic model shows ``us'' and the adversary's variables.
\\
{\bf Aim}: Referring to model (\ref{eq:model}) and  Figure \ref{fig:graph}, we address the following questions in this section:
\begin{compactenum} \item How to estimate the adversary's belief given measurements of its actions (which are based on  its filtered estimate of our state)? In other words, assuming  probability distributions  $\tp,\oprob,\aprob$ are  known\footnote{As mentioned in footnote $6$, this assumption simplifies the setup; otherwise we need to estimate the adversary's estimate of us, which makes our task substantially complex.}, we aim to estimate  the adversary's belief $\belief_k$ at each time $k$, by computing   posterior $\pdf(\belief_k\mid \belief_0,\state_{0:k},\action_{1:k})$. 
\item How to estimate the adversary's observation kernel $\oprob$, i.e its sensor gain? This  tells us how accurate the adversary's sensor is.
\end{compactenum}
From a practical point of view, estimating the adversary's belief and sensor parameters allows us to  calibrate its accuracy and predict (in a Bayesian sense)  future actions of the adversary.
\\ 
{\bf Related  Works}.
In recent works  \cite{MRKW17,MRKW18,MIC19}, the mapping from
belief $\belief$ to adversary's action $\laction$ was assumed  deterministic. In comparison, our proposed research here  assumes a probabilistic map between $\belief $ and $\action$ and we  develop Bayesian filtering algorithms for estimating  the posterior along with MLE (Maximum Likelihood Estimation) algorithms for estimating the underlying model. 
Estimating/reconstructing the posterior given decisions based on the posterior  is  studied in microeconomics under the area of
   social learning \cite{Cha04} and game-theoretic  generalizations 
\cite{AHP07}. 
There  are strong parallels between inverse filtering and Bayesian social learning  \cite{Cha04}, \cite{Kri16,Kri12,Kri11}; the key difference is that social learning aims to estimate the underlying state given noisy posteriors, whereas our aim is to estimate the posterior given noisy measurements of the posterior and the underlying state. Recently,  \cite{HAB18} used cascaded Kalman filters for LQG  control over  communication channels.  This work motivates the design of the function $\fun$ in \eqref{eq:linearaction} below that maps the adversary's belief to its action; see also footnote 5.
Finally, in \cite{CZ14}, the authors investigate the inverse problem of trajectory identification based on target measurements, where the target is assumed to follow a constant velocity model.

\subsection{Inverse Tracking Algorithms}\label{sec:inv_kalman}
\begin{quote}
{\bf {\em   How to estimate the adversary's posterior distribution  of us?}}
\end{quote}
Here we discuss inverse tracking for the model (\ref{eq:model}).
Define the posterior distribution $\post_{k}(\belief_k) = \pdf(\belief_k |\action_{1:k},\state_{0:k})$ of the adversary's posterior
distribution given our state sequence $\state_{0:k}$ and actions $\action_{1:k}$.
Note that the posterior $\post_k(\cdot)$ is a {\em random measure} since it is a posterior distribution of the adversary's posterior  distribution (belief) $\belief_k$.
By using a discrete time version of Girsanov's theorem and appropriate change of measure\footnote{This paper  deals with discrete time. Although we will not pursue it here, the recent paper \cite{KLS18} uses a similar  continuous time formulation. This  yields interesting results involving Malliavin derivatives  and stochastic calculus.} \cite{EAM95} (or  a careful application of Bayes rule) we can derive the following functional recursion for $\post_k$ (see \cite{KR19})
\beq
  \post_{k+1}(\belief) = \frac{\aprob_{\belief,\action_{k+1}}
    \,  \int_\Belief \oprob_{\state_{k+1}, \obs_{\belief_k,\belief}}\, \post_k(\belief_k) d\belief_k}
  {\int_\Belief \aprob_{\belief,\action_{k+1}}
    \,  \int_\Belief \oprob_{\state_{k+1}, \obs_{\belief_k,\belief}}\, \post_k(\belief_k) d\belief_k \, d\belief}
  \label{eq:post}
  \eeq
  Here $\obs_{\belief_k,\belief}$ is the observation such that $ \belief = \filter(\belief_k,\obs)$ where $\filter$ is the adversary's  filter (\ref{eq:belief}).  
  We call (\ref{eq:post})  the {\em optimal inverse filter} since it yields the Bayesian posterior of the adversary's  belief given our state and noisy measurements of the  adversary's actions.

\subsection*{Example: Inverse Kalman Filter} \label{sec:inversekalman}
We consider  a special case of (\ref{eq:post}) where the  inverse filtering problem  admits a finite dimensional characterization in terms of  the Kalman filter.
Consider a  linear Gaussian state space model
\beq \label{eq:lineargaussian}
\begin{split}
\state_{k+1} &= \statem\, \state_k  + \snoise_k, \quad \state_0 \sim \belief_0 \\
\obs_k &= \obsm\, \state_k + \onoise_k 
\end{split}
\eeq
where  $\state_k \in \statespace = \reals^\statedim$ is ``our'' state with
initial density $\belief_0 \sim \normal(\hat{\state}_0,\kalmancov_0)$,
 $\obs_k \in \obspace = \reals^\obsdim$ denotes the adversary's observations,
 $\snoise_k\sim \normal(0,\snoisecov_k)$,
 $\onoise_k \sim \normal(0,
\onoisecov_k)$
and 
  $\{\snoise_k\}$,  
  $\{\onoise_k\}$ are mutually independent  i.i.d.\ processes. Here, $\normal(\mu,C)$ denotes the normal distribution with mean $\mu$ and covariance matrix $C$.

 Based on observations $\obs_{1:k}$, the adversary computes the  belief  $\belief_k = \normal(\hstate_k,\kalmancov_k)$ where $\hstate_k$ is the conditional mean
  state   estimate and $\kalmancov_k$ is the covariance; these are computed via the classical Kalman filter
  equations:\footnote{For localization problems, we will use the information filter form:
    \beq  \kalmancov^{-1}_{k+1} = \kalmancov_{k+1|k}^{-1} + \obsm^\p \onoisecov^{-1} \obsm, \quad
\kg_{k+1} = \kalmancov_{k+1} \obsm^\p \onoisecov^{-1} \label{eq:info}
\eeq Similarly, the inverse Kalman filter in information form reads
\beq \enemykalmancov^{-1}_{k+1} = \enemykalmancov^{-1}_{k+1|k} +
\enemyobsm_{k+1}^\p \enemyonoisecov^{-1} \enemyobsm_{k+1},\;
\enemykg_{k+1} = \enemykalmancov_{k+1} \enemyobsm_{k+1}^\p \enemyonoisecov^{-1}. \label{eq:enemyinfo}
\eeq}
\beq
  \begin{split}
\kalmancov_{k+1|k} &=  \statem  \kalmancov_{k} \statem^\p  +  \snoisecov_k  \\
\Sig_{k+1} &= \obsm \kalmancov_{k+1|k} \obsm^\p + \onoisecov_k 
\\
{\hstate}_{k+1} &=  \statem\,  {\hstate}_k  + 
\kalmancov_{k+1|k} \obsm^{\p}  \Sig_{k+1}^{-1} 
(\obs_{k+1} - \obsm \, \statem\,  {\hstate}_k )
\\
\kalmancov_{k+1} &=
\kalmancov_{k+1|k} -  
\kalmancov_{k+1|k} \obsm^{\p}  \Sig_{k+1}^{-1} 
\obsm \kalmancov_{k+1|k} 
\end{split}
\label{eq:kalman}
\eeq
  The 
  adversary then chooses its  action as  $\eaction_k = \fun(\kalmancov_k)\,\hstate_k$ for some pre-specified function\footnote{In general the action $a_k$  is a function of the state estimate and covariance matrix. Choosing the action $a_k$ as a linear function of the  state estimate is for convenience and motivates the inverse Kalman filter discussed below. Moreover it  mimics linear quadratic Gaussian (LQG) control where the feedback is a linear function of the state estimate. In LQG control, the feedback gain is obtained from backward Riccati equation. Here we weigh by a nonlinear function of the Kalman covariance   matrix (forward Riccati equation)  to allow for incorporating uncertainty of the estimate into the choice of the action $a_k$.}
  $\fun$. We  measure the adversary's  action as
\beq \action_k = \fun(\kalmancov_k)\,\hstate_k + \anoise_k, \quad
\anoise_k \sim \text{ iid } \normal(0,\anoisecov) \label{eq:linearaction} \eeq

The Kalman covariance $\kalmancov_k$ is deterministic and fully determined by the model parameters. Hence, we only need to estimate  
$\hstate_k$ at each time $k$ given $a_{1:k},\state_{0:k}$ to estimate the belief $\belief_k=\normal(\hstate_k,\kalmancov_k)$.
Substituting (\ref{eq:lineargaussian})  for $\obs_{k+1}$ in (\ref{eq:kalman}), we see that
(\ref{eq:kalman}), (\ref{eq:linearaction}) constitute a linear Gaussian system with unobserved state  $\hstate_k$, observations $\action_k$,
and known exogenous  input $\state_k$:
\beq \label{eq:inversekf}
\begin{split}
  {\hstate}_{k+1} &=   (\vect I - \kg_{k+1} \obsm) \, \statem \hstate_{k} + \kg_{k+1} \onoise_{k+1} + \kg_{k+1} \obsm \state_{k+1} \\
   \action_k &= \fun(\kalmancov_k)\,\hstate_k + \anoise_k, \quad
   \anoise_k \sim \text{ iid } \normal(0,\anoisecov), \\
  & \text{ where }  \kg_{k+1} = \kalmancov_{k+1|k} \obsm^{\p}  \Sig_{k+1}^{-1}. 
\end{split}
\eeq
$\kg_{k+1}$ is called the Kalman gain and $\vect I$ is the identity matrix.

To summarize,  our filtered estimate of the adversary's filtered estimate    $\hstate_k$ given measurements $a_{1:k},\state_{0:k}$ is achieved by running ``our''  Kalman filter on the linear Gaussian state space model  (\ref{eq:inversekf}), where
$\hstate_k, \kg_k, \kalmancov_k$ in (\ref{eq:inversekf}) are generated by the adversary's  
Kalman filter. Therefore, our Kalman filter uses the parameters
\beq
\begin{split}
\enemystatem_{k}  &=  (\vect I - \kg_{k+1} \obsm)\statem, \;
\enemyinputm_k = \kg_{k+1} \obsm,\;
  \enemyobsm_k =  \fun(\kalmancov_k), \\
\enemysnoisecov_k  &= \kg_{k+1}\, \kg_{k+1}^\p, \;
\enemyonoisecov_k = \anoisecov
\end{split} \label{eq:inversekfparam}
  \eeq
 The equations of our inverse Kalman filter for estimating the adversary's estimate of our state are:
\beq
 \begin{split}
  \enemykalmancov_{k+1|k} &=  \enemystatem_k \enemykalmancov_{k} \enemystatem_k^\p  +  \enemysnoisecov_k  \\
  \enemySig_{k+1} &= \enemyobsm_{k+1} \enemykalmancov_{k+1|k} \enemyobsm_{k+1}^\p + \enemyonoisecov_k  \\
   \enemystate_{k+1} &= \enemystatem_k\, \enemystate_k + 
\enemykalmancov_{k+1|k} \enemyobsm_{k+1}^{\p}  \enemySig_{k+1}^{-1} 
                        \\ & \pushright{\times \big[\action_{k+1} -   \enemyobsm_{k+1} \left(\enemystatem_{k} \enemystate_k+ \enemyinputm_k \state_{k+1} \right) \big]} 
  \\
  & \pushright{ + \enemyinputm_k \state_{k+1} }\\
   \enemykalmancov_{k+1} &=
\enemykalmancov_{k+1|k} -  
\enemykalmancov_{k+1|k} \enemyobsm_{k+1}^{\p}  \enemySig_{k+1}^{-1} 
\enemyobsm_{k+1} \enemykalmancov_{k+1|k} 
\end{split} \label{eq:inversekfequations}
\eeq
Note $\enemystate_k$ and $\enemykalmancov_k$ denote our conditional mean estimate and covariance of the adversary's conditional mean $\hstate_k$.
The computational cost of the inverse Kalman filter is identical to the classical Kalman filter, namely $O(\statedim^2)$ computations at each time step.

  {\bf Remarks}:   
  \begin{compactenum}
\item  As discussed in \cite{KR19},  inverse Hidden Markov model (HMM) filters and inverse particle filters can also  be derived to solve the inverse tracking problem. For example, the inverse HMM filter  deals with the case 
    when $\belief_k$ is computed via an HMM filter and the estimates of the HMM filter are observed in noise. In this case the inverse filter has a computational cost that grows exponentially with the size of the observation alphabet.
\item A general approximate solution for (\ref{eq:post}) involves sequential Markov chain Monte-Carlo (particle filtering).
  In particle filtering, cases where it is possible to sample from the so called optimal importance function are of significant interest \cite{RAG04,CMR05}.
  In inverse filtering, \cite{KR19}  shows  that the optimal importance function can be determined explicitly due to the structure of the inverse filtering problem. Specifically, in our case, the ``optimal'' importance density is  $\imp^* = \pdf(\belief_k,\obs_k| \belief_{k-1},\obs_{k-1}, \state_{k},\action_{k} )$.
Note that in our case 
\begin{align}
\imp^*=~& \pdf(\belief_k | \belief_{k-1},\obs_k)\, \pdf(\obs_k| \state_k,\action_k)\nonumber\\
=~& \delta\big(\belief_k - \filter(\belief_{k-1},\obs_k)\big)\, \pdf(\obs_k|\state_k)\label{eq:imp_density}
\end{align}
is straightforward to sample from. 
There has been a substantial amount of recent research in finite sample concentration bounds for the particle filter \cite{DEL11,MAR18}. In future work such results can be used to evaluate the  sample complexity of the inverse particle filter.
  
\end{compactenum}

\subsection{Estimating the Adversary's Sensor Gain}
\label{sec:adv_gain}
 In this section, we discuss how to  estimate the adversary's sensor observation kernel  $\oprob$ in (\ref{eq:model}) which  quantifies the accuracy of  the adversary's sensors.
We assume that  $\oprob$   is parameterized by an $\modeldim$-dimensional vector $\model \in \Model$ where
$\Model$ is a compact subset of $\reals^\modeldim$. Denote the parameterized
observation kernel as $\oprob^\model$.
Assume  that both us and the adversary know \footnote{Otherwise the adversary estimates $\tp$ as $\hat{\tp}$ and we need to estimate the adversary's estimate of us, namely 
$\hat{\hat{\tp}}$.  This makes the estimation task substantially more complex. In future work we will examine conditions under which the MLE in this setup is identifiable  and consistent.}) $\tp$ (state transition kernel and $\aprob$ (probabilistic map from adversary's belief to its action). As mentioned earlier, the stochastic kernel $\aprob$ in \eqref{eq:model} is a composition of two stochastic kernels: 1) the map from the adversary's belief $\belief_k$ to its action $\laction_k$, and 2) the map from the adversary's action $\laction_k$ to our measurement $\action_k$ of this action.

Then, given our state sequence $\state_{0:\horizon}$  and adversary's action sequence $\laction_{1:\horizon}$, our aim is to compute the maximum likelihood estimate (MLE) of $\model$. That is, with $\lik(\model)$ denoting the log-likelihood, the aim is to compute
\beq \label{eqn:MLE_def} \mle = \argmax_{\model \in \Model} \lik(\model),~\lik(\model)= \log \pdf(\state_{0:\horizon},\action_{1:\horizon} | \model) .\eeq
The likelihood can be evaluated from
 the un-normalized  inverse filtering recursion (\ref{eq:post}) 
 \begin{align}   
 \lik(\model) &= \log \int_\Belief \unpostm_\horizon(\belief) d\belief,\nonumber\\
\unpostm_{k+1}(\belief) &= \aprob_{\belief,\action_{k+1}}
    \,  \int_\Belief \oprob^\model_{\state_{k+1}, \obs^\model_{\belief_k,\belief}}\, \unpostm_k(\belief_k) d\belief_k,
\label{eq:unpost}
\end{align}
initialized by setting $\unpostm_0(\belief_0) = \belief_0$. Here $\obs^{\model}_{\belief_k,\belief}$ is the observation such that $ \belief = \filter(\belief_k,\obs)$ where $\filter$ is the adversary's  filter (\ref{eq:belief}) with variable $\oprob$ parametrized by $\model$.
Given (\ref{eq:unpost}), a local stationary point of the likelihood can be computed using  a general purpose numerical optimization algorithm.


\subsection{Example. Estimating Adversary's Gain in Linear Gaussian case}
The aim of this section is to provide insight into the nature of estimating the adversary's sensor gain via numerical examples.
Consider the setup  in Sec.\ref{sec:inversekalman} where our dynamics are linear Gaussian and the adversary observes our state linearly in Gaussian noise (\ref{eq:lineargaussian}). The adversary
estimates our state using a Kalman filter, and we estimate the adversary's estimate using the inverse Kalman filter (\ref{eq:inversekf}).
Using  (\ref{eq:inversekf}), (\ref{eq:inversekfparam}),  the log-likelihood  for the adversary's observation  gain matrix $\param=\hobsm$  based on our measurements is\footnote{The variable $\param$ is introduced only for notational clarity.}
\begin{align}
  \lik(\model) &= \text{const} - \frac{1}{2} \sum_{k=1}^\finaltime \log | \enemySig^\param_k| - \frac{1}{2} \sum_{k=1}^\finaltime
  \innovations_k^\p\, (\enemySig_k^\param)^{-1} \,\innovations_k\nonumber \\
  \innovations_k &= \action_k - \enemyobsm_k^\param\, \enemystatem_{k-1}^\param \enemystate_{k-1} - \enemyinputm_{k-1}^\param \state_{k-1} \label{eq:inversekflik}
\end{align}
where $\innovations_k$ are the innovations of the inverse Kalman filter (\ref{eq:inversekfequations}).
In (\ref{eq:inversekflik}), our state $\state_{k-1}$ is known to us and therefore is a known exogenous input. Also note from (\ref{eq:inversekfparam}) that $\enemystatem_{k}, \enemyinputm_k$
are explicit functions of $\obsm$, while $\enemyobsm_k$ and $\enemysnoisecov_k$ depend on $\obsm$ via the adversary's Kalman filter.

\begin{figure}[h]
\includegraphics[width=0.48\textwidth]{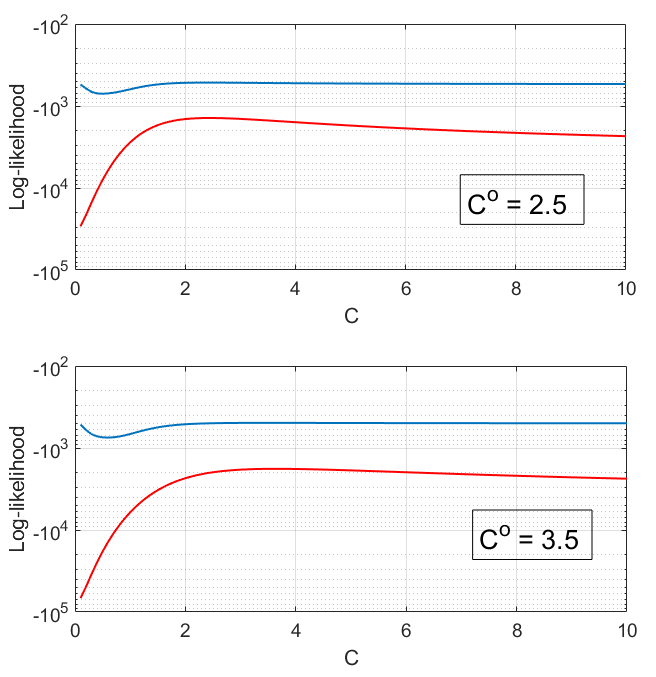}
\caption{Log-Likelihood as a function of adversary's gain $\hobsm\in (0,10]$ when true value is $C^o$. The red curves denote the log-likelihood of $\hobsm$ given the adversary's measurements of our state. The blue curves denote the log-likelihood of $\hobsm$ using the inverse Kalman filter given our observations of the adversary's action $\laction_k$. The plots show that it is more difficult to compute the MLE \eqref{eqn:MLE_def} for the inverse filtering problem due to the almost flat likelihood (blue curves) compared to red curves.}
\label{fig:kf}
\end{figure}

 The log-likelihood  for the adversary's observation  gain matrix $\param=\hobsm$  can be evaluated using (\ref{eq:inversekflik}).
To provide insight,  Figure \ref{fig:kf} displays  the log-likelihood versus adversary's gain matrix $\hobsm$  in the scalar case for 1000 equally spaced data points over the interval $\obsm=(0,10]$.
 The four sub-figures correspond to true values $\obsmt = 2.5,3.5$ of $\obsm$, respectively.

Each sub-figure in  Figure \ref{fig:kf} has two plots. The plot in red is  the  log-likelihood of $\hat{\obsm} \in (0,10]$ evaluated based on the adversary's observations using the  standard  Kalman filter. (This is the classical log-likelihood of the observation gain of a Gaussian state space model.)  The plot in blue is the log-likelihood of $\hobsm\in (0,10]$ computed using  our measurements of the adversary's action using the  inverse Kalman filter (where the adversary first estimates our state using a Kalman filter) - we call this the inverse case.

 Figure \ref{fig:kf} shows  that the log-likelihood in the inverse case (blue plots) has a less pronounced maximum compared to the standard case (red plots). Therefore, numerical algorithms for computing the MLE of the  adversary's gain $\obsmt$ using our observations of the adversary's actions (via the inverse Kalman filter) will converge much more slowly than the classical MLE (based on  the adversary's observations). This is intuitive  since  our estimate of the adversary's parameter is based on the adversary's estimate of our state and so has more noise.

{\bf Sensitivity of  MLE.} 
It is important to evaluate the sensitivity of the MLE of $\obsm$ wrt covariance matrices $\snoisecov_k,~\onoisecov_k$ in the state space model~\eqref{eq:lineargaussian}. For example,
the sensitivity wrt $\snoisecov_k$ reveals how sensitive the MLE is wrt our maneuver covariance  since from
\eqref{eq:lineargaussian}, $\snoisecov_k$ determines our maneuvers.
Our sensitivity analysis evaluates the variation of the second derivative of the log-likelihood of $\obsm$ computed at the true gain $\obsm^o$ to small changes in $\snoisecov_k$ and $\onoisecov_k$. Table \ref{tab:sensitivity} displays our sensitivity results wrt the scalar setup of Figure \ref{fig:kf}. Table~\ref{tab:sensitivity} comprises two sensitivity values,
\beq \label{eq:sensitivity}
\begin{split}
 \sense_{\snoisecov}=&\frac{\partial}{\partial \snoisecov_k} \left( \frac{\partial^2 \lik(\model)}{\partial \model^2} \right)\bigg\vert_{\model=\obsm^o}~\text{and}\\
 \sense_{\onoisecov}=&\frac{\partial}{\partial \onoisecov_k} \left( \frac{\partial^2 \lik(\model)}{\partial \model^2} \right)\bigg\vert_{\model=\obsm^o},
 \end{split}
\eeq
evaluated for both the inverse case (that uses the inverse Kalman filter \eqref{eq:inversekflik}) and the classic case where the adversary's observations are known. $\sense_{(\cdot)}$ measures the change in the sharpness of the log-likelihood plot around the true sensor gain wrt change in the noise covariance. 
Note that the experimental setup of Figure \ref{fig:kf} assumes the covariances $\snoisecov_k,\onoisecov_k$ are constant over time index $k$, hence we drop the subscript in the LHS of \eqref{eq:sensitivity}. 

Table \ref{tab:sensitivity} shows that the second derivative of the log-likelihood is more sensitive (in magnitude) to the adversary's observation covariance $\onoisecov_k$ than the maneuver covariance $\snoisecov_k$. Also, it is observed that the sensitivity of the log-likelihood is higher for lower sensor gain $\obsm^o$. This observation is consistent with intuition since a larger gain $\obsm$ implies a larger SNR (signal-to-noise ratio) of the observation $\obs_k$ which intuitively suggests the estimate of $\obsm$ is more robust to changes in maneuver covariance and observation noise covariance.

\begin{table} \centering
\begin{tabular}{ |l|l|l|l| }
\hline
 & $\obsm^o$ & Classic & Inverse\\ \hline
\multirow{2}{*}{$\sense_{\snoisecov}$} & $2.5$ & $-43.45$ & $-6.46$ \\ 
 & $3.5$ & $-25.16$ & $-2.77$\\ \hline
\multirow{2}{*}{$\sense_{\onoisecov}$} & $2.5$ & $-189.39$ & $-50.04$ \\ 
 & $3.5$ & $-65.27$ & $-30.55$ \\ \hline
\end{tabular}
\caption{Comparison of sensitivity values \eqref{eq:sensitivity} for log-likelihood of $\obsm$ wrt noise covariances $\snoisecov_k,~\onoisecov_k$ \eqref{eq:lineargaussian} - classical  model vs inverse Kalman filter model.}
\label{tab:sensitivity}
\end{table}

{\bf Cram\'er-Rao (CR) bounds}. It is instructive to compare the   CR bounds for MLE of $\obsm$ for the classic model versus that of the inverse Kalman filter model. Table \ref{tab:cr} displays the CR bounds (reciprocal of expected Fisher information) for the four examples considered above evaluated using via the algorithm in  \cite{CS96}. 
It shows that the covariance lower bound for the inverse case is substantially higher than that for the classic case. This is
consistent with the intuition that estimating the adversary's parameter based on its actions (which is based on its estimate of us) is more difficult than directly estimating $\obsm$ in a classical state space model based on the adversary's observations of our state that determines its actions.

\begin{table} \centering
\begin{tabular}{|c|c|c|}
  \hline
  $\obsmt$ & Classic & Inverse \\
  \hline
  0.5 & $0.24 \times 10^{-3} $ & $5.3 \times 10^{-3}$ \\
  1.5 & $1.2 \times 10^{-3}$ &   $37 \times 10^{-3}$ \\
  2    & $2.1 \times 10^{-3} $ &  $70 \times 10^{-3}$ \\
  3  & $ 4.6 \times 10^{-3}$ & $ 336 \times 10^{-3}$\\
  \hline                       
\end{tabular}
\caption{Comparison of Cram\'er-Rao bounds for $\obsm$ - classical  model vs inverse Kalman filter model.}
\label{tab:cr}
\end{table}

{\em Consistency of MLE}. The above example (Figure~\ref{fig:kf}) shows that the likelihood surface
of  $ \lik(\model)= \log \pdf(\state_{0:\horizon},\action_{1:\horizon} | \model)$ is flat and hence computing the MLE numerically can be difficult.  Even in the case when we observe the adversary's actions perfectly, \cite{MRKW17} shows that non-trivial  observability conditions need to be imposed on the system parameters.
  
For the linear Gaussian case where we observe the adversary's Kalman filter in noise,
strong consistency  of the MLE for the adversary's gain matrix $\obsm$ can be established
fairly straightforwardly. 
Specifically, if we 
assume that state matrix $\statem$ is stable, and the state space model 
is an identifiable minimal realization, then the adversary's Kalman filter variables 
 converge to steady state values geometrically fast in  $k$  \cite{AM79} implying that asymptotically the inverse Kalman filter system  is stable linear time invariant. Then, the  MLE $\mle$  for the adversary's observation matrix $\obsm$   is unique and strongly consistent \cite{Cai88}.


\section{Identifying Utility Maximization in a Cognitive Radar}\label{sec:RP_UM}

The previous section was concerned with estimating the adversary's posterior belief and sensor accuracy.  
This section discusses detecting utility maximization behavior and estimating the adversary's utility function
in the context of cognitive radars. As described in the introduction, inverse tracking, identifying utility maximization and designing interference to confuse the radar constitute our adversarial setting.

Cognitive radars \cite{Hay06} use  the perception-action cycle of cognition to  sense the environment and learn from it relevant information about the target and the environment. The cognitive radars then tune the radar sensor to optimally satisfy their mission objectives.
Based on its tracked estimates, the cognitive radar adaptively optimizes its waveform, aperture, dwell time and revisit rate. In other words, a cognitive radar is a constrained utility maximizer.

This section   is motivated by the  next logical step, namely, \textit{identifying a cognitive radar} from the actions of the radar.
 The adversary cognitive radar
observes our  state in noise; it  uses a Bayesian estimator (target tracking algorithm)  to update its posterior
distribution of our state and then chooses an action based on this
posterior.
From the intercepted emissions of an adversary's radar, we address the following question: 
    Are the adversary sensor's actions consistent with optimizing a monotone utility function (i.e., is the cognitive sensor behavior rational in an economics sense)? If so how can we estimate a utility function of the adversary's  cognitive sensor that is consistent with its actions?
The main synthesis/analysis framework we will use   is  that of revealed preferences \cite{Var12,FM09,Die12} from microeconomics which
aims to determine preferences by observing choices. The results presented below are developed in detail in the recent work \cite{KAEM20}; however, the SINR constraint formulation in Sec.\,\ref{sec:waveform}  for detecting waveform optimization is new. Related work that develops adversarial inference strategies at a higher level of abstraction than tracking level (like the Bayesian filter level in Sec.\,II) includes \cite{Kup17}. \cite{Kup17} places counter unmanned autonomous systems at a level of abstraction above the physical sensors/actuators/weapons and datalink layers; and below the human controller layer.

    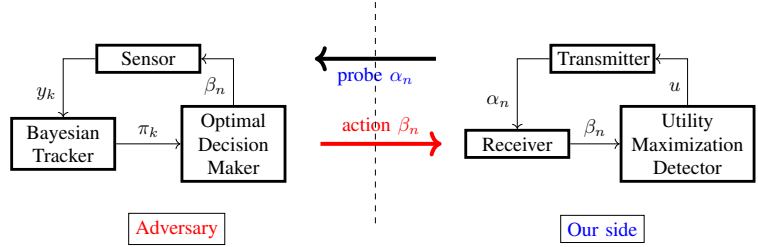
\begin{figure*}
    \centering
            {\resizebox{10cm}{!}{
\begin{tikzpicture}[node distance = 1cm, auto]
    \node [blocka] (BLOCK1) {Sensor};
    \node [blocka, below of=BLOCK1,right of=BLOCK1,node distance=1.5cm] (BLOCK2) {Optimal Decision \\ Maker};
    \node [blocka, below of=BLOCK1,left of=BLOCK1,node distance=1.5cm] (BLOCK3) {Bayesian Tracker};


    
    \node [blocka, right of = BLOCK1, node distance = 8cm] (BLOCK4) {Transmitter};
    \node [blockb, below of=BLOCK4,right of=BLOCK4,node distance=1.5cm] (BLOCK5) {Utility Maximization Detector} ;
    \node [blocka, below of=BLOCK4,left of=BLOCK4,node distance=1.5cm] (BLOCK6) {Receiver} ;
    
    \draw[<-] (BLOCK4) -| node[left,pos=0.8]{$u$}  (BLOCK5)  ;
    \draw[->] (BLOCK4.west) -|   node[left,pos=0.8]{$\probe_\dtime$} (BLOCK6);

    \draw[->](BLOCK6) --  node[above]{$\response_\dtime$} (BLOCK5);

    \draw[<-] (BLOCK1) -| node[left,pos=0.8]{$\response_\dtime$}  (BLOCK2)  ;
    \draw[->] (BLOCK1.west) -|   node[left,pos=0.8]{$\obs_k$} (BLOCK3);

    \draw[->](BLOCK3) --  node[above]{$\belief_k$} (BLOCK2);


    \draw[->,color=red,line width=2pt] ([xshift=0.6cm]BLOCK2.east)   -- node[above]{action $\response_\dtime$} ([xshift=-0.4cm]BLOCK6.west);
    \draw[->,line width=2pt] ([xshift=-2cm]BLOCK4.west)   --   node[below]{\textcolor{blue}{probe $\probe_\dtime$}} ([xshift=2cm]BLOCK1.east);
    \node[draw] at (8,-3.0) {{\color{blue}Our side}};
    \node[draw] at (0.5,-3.0) {{\color{red}Adversary}};
    \draw [dashed] (4,1) -- (4,-3);
   \end{tikzpicture}} }

\caption{Schematic of Adversarial Inference Problem. Our side is a drone/UAV or electromagnetic signal that probes the  adversary's cognitive radar system. $k$ denotes a fast time scale and $n$ denotes a slow time scale. Our state $\state_k$, parameterized by $\probe_\dtime$ (purposeful acceleration maneuvers), probes the adversary radar. Based on the noisy observation $\obs_k$ of our state, the adversary radar  responds with action~$\response_\dtime$. Our aim (in the Utility Maximization Detector block) is to detect if the adversary radar is economic rational, i.e.,  is its response  $\beta_\dtime$  generated by constrained optimizing a utility function $u$, and if so, estimate the utility function.}
\label{fig:schematic}
\end{figure*}

\subsection{Background. Revealed Preferences and Afriat's Theorem} Non-parametric detection of utility maximization behavior is studied in the area of revealed preferences in microeconomics.
A key result is the following:
\begin{definition}[\cite{Afr67,Afr87}]\label{eq:utility_maximizer}
A system is  a {\em utility maximizer} if for every probe $\probe_\dtime\in \reals_+^\probedim$, the  response $\response_\dtime \in \reals^\probedim$ satisfies
\begin{equation}
  \response_\dtime\in \argmax_{\probe_\dtime^\p \response \leq 1}\utility(\response) 
\label{eq:utilitymaximization}
\end{equation}
where $\utility(\response)$ is a {\em monotone} utility function.
 \end{definition}

In economics, $\probe_\dtime$ is the price vector and  $\response_\dtime$ the consumption vector. Then $\probe_\dtime^\p \response \leq 1$ is a natural budget constraint\footnote{The budget constraint $\probe_\dtime^\p \response \leq 1$ is without loss of generality, and can be replaced by $\probe_\dtime^\p \response \leq c$ for any positive constant $c$. A more general nonlinear budget incorporating spectral constraints will be discussed  below.}  for a consumer with 1 dollar. Given a dataset of price and consumption vectors, the aim is to determine
   if the consumer is a utility maximizer (rational) in the sense of (\ref{eq:utilitymaximization}).  
 
 The key result  is the 
following  theorem  
due to Afriat~\cite{Die12,Afr87,Afr67,Var12,Var83}

\begin{theorem}[Afriat's Theorem~\cite{Afr67}] Given a data set
$
 \dataset=\{(\probe_\dtime,\response_\dtime), \dtime\in \{1,2,\dots,\horizon\}\}$,
 the following statements are equivalent:
	\begin{compactenum}
	\item The system is a utility maximizer and there exists a monotonically increasing, continuous,  and concave utility function that satisfies (\ref{eq:utilitymaximization}). 
		\item There exist positive reals $u_t,\lagrange_t>0,~t=1,2,\ldots,\horizon,$ such that the following inequalities hold.
			\begin{equation}
				u_\sindx-u_\tindx-\lambda_\tindx \probe_\tindx^\p (\response_\sindx-\response_\tindx) \leq 0 \; \forall \tindx,\sindx\in\{1,2,\dots,\horizon\}.\
				\label{eqn:AfriatFeasibilityTest}
			\end{equation}
			The monotone, concave utility function\footnote{As pointed out in \cite{Var82}, a remarkable feature of Afriat's theorem is that if the dataset can be rationalized by a monotone utility function, then it can be rationalized by a continuous, concave, monotonic utility function. Put another way,   continuity and  concavity cannot be refuted with  a finite datasaset.} given by 
			\begin{equation}
				\utility(\response) = \underset{\tindx\in \{1,2,\dots,\horizon\}}{\operatorname{min}}\{u_\tindx+\lambda_\tindx \probe_\tindx^\p(\response-\response_\tindx)\}
				\label{eqn:estutility}
            \end{equation}
            constructed using $u_t$ and $\lagrange_t$ defined in \eqref{eqn:AfriatFeasibilityTest}            rationalizes the dataset  by satisfying \eqref{eq:utilitymaximization}.
			
          \item The data set $\mathcal{D}$ satisfies the Generalized Axiom of Revealed Preference (GARP) also called cyclic consistency, namely for any $\tindx \leq \horizon$,
           $\probe_t^\p \response_t \geq \probe_t^\p \response_{t+1} \quad \forall t\leq k-1 \implies \probe_k^\p  \response_k \leq \probe_k^\p  \response_{1}$.
	\end{compactenum}
\label{thm:AfriatTheorem}
\end{theorem}

Afriat's theorem tests for economics-based rationality; its  remarkable property is that it gives a {\em necessary and sufficient condition} for a system  to be a utility maximizer based on the system's input-output response. Although GARP in statement 4 in Theorem~\ref{thm:AfriatTheorem} is not critical to the developments in this paper, it is of high significance in micro-economic theory and is stated here for completeness.
The feasibility of the set of inequalities (\ref{eqn:AfriatFeasibilityTest}) can be checked using a linear programming solver; alternatively GARP   can be checked  using Warshall's algorithm with $O(\horizon^3)$ computations~\cite{Var06,Var82}.

The recovered utility using~\eqref{eqn:estutility}  is not unique; indeed  any positive monotone increasing transformation of~\eqref{eqn:estutility} also satisfies Afriat's Theorem; that is, the utility function constructed is ordinal. This is the reason why the budget constraint $\probe_\dtime^\p \response \leq 1$ is without generality; it can be scaled by an arbitrary positive  constant and  Theorem \ref{thm:AfriatTheorem} still holds.  In signal processing terminology, Afriat's Theorem can be viewed as set-valued system identification of an \emph{argmax} system; set-valued since (\ref{eqn:estutility}) yields a set of utility functions that rationalize the finite dataset $\dataset$.

\subsection{Beam Allocation: Revealed Preference Test}\label{sec:beam}
This section constructs a test to identify a  cognitive
radar that switches its beam adaptively between targets.  This example is based on~\cite{KAEM20} and is presented here for completeness
The setup is schematically shown in Figure~\ref{fig:schematic}. We view each component $i$  of the probe signal $\probe_\dtime(i)$ as the trace of the information matrix (inverse covariance) of target $i$. We  use the trace of the information matrix of each target in our probe signal -- this allows us to consider multiple targets. Since the adversarial radar is assumed to be stationary, the target covariance used to define the probe for the radar is indeed the maneuver covariance.

The setup in Figure \ref{fig:schematic} differs significantly from the setup of Figure \ref{fig:graph} considered in the previous section. First, the adversary in the current setup is an economically rational agent. In Figure \ref{fig:graph}, the adversary is only specified at a lower level of abstraction as using a Bayesian filter to track our maneuvers. Second, this section abstracts adversary's actions at the fast time scale indexed by $k$ by an appropriately defined response at the slow time scale indexed by $n$. The previous section's analysis was confined to the actions generated only at the fast time scale $k$. Lastly, Figure \ref{fig:schematic} assumes the abstracted response $\response_k$ of the adversary is measured accurately by us as opposed to a noisy measurement $\action_k$ of the adversary's action $\laction_k$ in Figure \ref{fig:graph}.

Suppose a radar adaptively switches its beam between
$\probedim$ targets where these $\probedim$ targets are controlled by us. As in (\ref{eq:lineargaussian}), on the fast time scale indexed by $k$, each target $i$ has linear Gaussian dynamics  and the adversary  radar obtains linear Gaussian measurements:
\beq \label{eq:lineargaussian2}
\begin{split}
\state^i_{k+1} &= \statem\, \state^i_k  + \snoise^i_k, \quad \state_0 \sim \belief_0 \\
\obs^i_k &= \obsm\, \state^i_k + \onoise^i_k, \quad i=1,2,\ldots,\probedim
\end{split}
\eeq
Here $ \snoise^i_k\sim \normal(0,\snoisecov_\dtime(i))$,
 $\onoise_k^i \sim \normal(0,
 \onoisecov_\dtime(i))$. Recall from Figure \ref{fig:schematic} that $n$ indexes the epoch (slow time scale) and $k$ indexes the fast time scale within the epoch.
 We assume that both $\snoisecov_\dtime(i)$ and $\onoisecov_\dtime(i)$ are known to us and the adversary.

 The adversary's radar tracks our $\probedim$ targets using  Kalman filter trackers.
 The fraction of time the radar allocates to each target $i$ in epoch $\dtime$ is $\response_\dtime(i)$. 
The price the radar pays for each target $i$ at the beginning of epoch $\dtime$ is the trace of the  predicted {\em accuracy} of target $i$. Recall that this is  the trace
of the inverse of the predicted  covariance   at  epoch $\dtime$ using the Kalman predictor
\beq  \probe_\dtime(i) = \trace(\kalmancov^{-1}_{\dtime|\dtime-1}(i)), \quad i =1,\ldots, \probedim
\label{eq:probe_beam}
\eeq
The predicted covariance $\kalmancov_{\dtime|\dtime-1}(i)$ is a deterministic function of  the maneuver covariance $\snoisecov_\dtime(i)$ of target $i$.  So the probe\footnote{In comparison to \eqref{eq:lineargaussian}, the velocity and acceleration elements of $\state_k^i$ in \eqref{eq:lineargaussian2} must be multiplied by normalization factors $\Delta t$ and $(\Delta t)^2$ respectively, for \eqref{eq:probe_beam} to be dimensionally correct, where $\Delta t$ is the time duration between two discrete time instants on the fast time scale.} $\probe_\dtime(i)$ is a signal that we can choose, since it is a deterministic function of  the maneuver covariance $\snoisecov_\dtime(i)$ of target $i$. We abstract the target's covariance by its  trace denoted by $\probe_\dtime(i)$.  Note also that  the observation noise covariance $\onoisecov_\dtime(i)$ depends on the adversary's radar response $\response_\dtime(i)$, i.e.,  the fraction of time allocated to target $i$.
We assume that each target $i$ can estimate the fraction of time $\response_\dtime(i)$ the adversary's radar allocates to it using a radar detector.

Given the time series $\probe_\dtime, \response_\dtime$, $\dtime = 1,\ldots,\horizon$,
our aim is to detect if the adversary's radar is cognitive. We assume that 
a cognitive radar optimizes its  beam allocation as the following constrained optimization:
\begin{equation}
  \begin{split}
    \response_\dtime &=     \argmax_\response ~\utility(\response)  \\
  \text{ s.t. } &   \response^\p \probe_\dtime \leq \precconstraint, 
  \end{split} \label{eq:beam}
\end{equation}
where $\utility(\cdot)$ is the adversary radar's utility function (unknown to us)  and $\precconstraint \in \reals_+$ is a pre-specified average accuracy of  all $\probedim$  targets.

The economics-based rationale  for the budget constraint is natural: For targets that are cheaper (lower accuracy $\probe_\dtime(i)$), the radar  has incentive to  devote more time $\response_\dtime(i)$. However, given its resource constraints,
the radar can  achieve at most an  average accuracy of $\precconstraint$ over all targets.

The setup in (\ref{eq:beam}) is directly amenable  to   Afriat's Theorem \ref{thm:AfriatTheorem}.
Thus  (\ref{eqn:AfriatFeasibilityTest}) can be used to test if the radar satisfies utility maximization in its beam scheduling (\ref{eq:beam}) and also estimate the set of utility functions~(\ref{eqn:estutility}).
Furthermore (as in Afriat's theorem) since the utility is ordinal, $\precconstraint$ can be chosen as 1 without loss of generality (and therefore does not need to be known by us).  

\subsection{Waveform adaptation: Revealed Preference Test for Non-linear budgets}\label{sec:waveform}
In the previous subsection, we tested for cognitivity of a radar by viewing it as an abstract system that switches its beam adaptively between targets. Here, we discuss cognitivity with respect to waveform design. Specifically, we construct a test to identify  cognitive behavior of an adversary radar that optimizes its waveform based on the SINR of the target measurement. By using a generalization of Afriat's theorem~(Theorem~\ref{thm:AfriatTheorem}) to non-linear budgets, our main aim is to detect if a radar intelligently chooses its waveform to maximize an underlying utility subject to signal processing constraints. Our setup below differs from \cite{KAEM20} since we introduce the SINR as a nonlinear budget constraint; in comparison \cite{KAEM20} uses a spectral budget constraint.

We start by briefly outlining the generalized utility maximization setup. 
\begin{definition}[\cite{FM09}]\label{eq:utility_maximizer_nonlinear}
A system is  a {\em generalized utility maximizer} if for every probe $\probe_\dtime\in \reals_+^\probedim$, the  response $\response_\dtime \in \reals^\probedim$ satisfies
\begin{equation}
  \response_\dtime\in \argmax_{g_n(\response) \leq 0}\utility(\response) 
\label{eq:utilitymaximization_nonlinear}
\end{equation}
where $\utility(\response)$ is a {\em monotone} utility function and $g_n(\cdot)$ is monotonically increasing in $\beta$.
\end{definition}
The above utility maximization model generalizes Definition~\ref{eq:utility_maximizer} since the budget constraint $g_n(\beta)\leq 0$ can accommodate non-linear budgets and includes the linear budget constraint of Definition~\ref{eq:utility_maximizer} as a special case. The result below provides an explicit test for a system that maximizes utility in the sense of Definition~\ref{eq:utility_maximizer_nonlinear} and constructs a set of utility functions that rationalizes the decisions $\beta_n$ of the utility maximizer. 
\begin{theorem}[Test for rationality with nonlinear budget~\cite{FM09}] Let $B_n=\{\beta\in\reals_{+}^m|g_n(\beta)\leq 0\}$ with $g_n:\reals^m\rightarrow\reals$ an increasing, continuous function and $g_n(\beta_n)=0$ for $n=1,\ldots N$. Then the following conditions are equivalent:
\begin{compactenum}[1)]
\item There exists a monotone continuous utility function $U$ that
rationalizes the data set $\{\beta_n, B_n\},n=1,\ldots N$. That is
\begin{equation*}
    \beta_n = \argmax_{\beta} U(\beta),\quad g_n(\beta)\leq 0
\end{equation*}
\item There exist positive reals $u_t,\lagrange_t>0,~t=1,2,\ldots,\horizon,$ such that the following inequalities hold.
\begin{equation}\label{eqn:lagrange_nonlinearafriat}
    u_s-u_t-\lambda_t g_t(\beta_s)\leq 0\quad \forall t,s\in\{1,2,\ldots N\}
\end{equation}
The monotone, concave utility function given by
\begin{equation}\label{eqn:construct_nonlinearafriat}
    U(\beta) = \min_{t\in\{1,\ldots N\}} \{u_t + \lambda_t g_t(\beta)\}
\end{equation}
constructed using $u_t$ and $\lambda_t$ defined in~(\ref{eqn:lagrange_nonlinearafriat}) rationalizes the data set by satisfying \eqref{eq:utilitymaximization_nonlinear}.
\item The data set $\{\beta_n,B_n\}, n = 1,\ldots, N$ satisfies $GARP$:
\begin{equation}\label{eqn:GARP_nonlinear}
    g_t(\beta_j)\leq g_t(\beta_t) \implies g_j(\beta_t)\geq 0
\end{equation}
\end{compactenum}
\label{thm:nonlinearAfriat}
\end{theorem}
Like Afriat's theorem, the above result provides a {\em necessary and sufficient condition} for a system to be a utility maximizer based on the system's input-output response. In spite of a non-linear budget constraint, it can be easily verified that the constructed utility function $U(\beta)$~(\ref{eqn:construct_nonlinearafriat}) is ordinal since any positive monotone increasing transformation of~(\ref{eqn:construct_nonlinearafriat}) satisfies the GARP inequalities~(\ref{eqn:GARP_nonlinear}).  

We now justify the non-linear budget constraint in~(\ref{eq:utilitymaximization_nonlinear}) in the context of the cognitive radar by formulating an optimization problem the radar solves equivalent to Definition~\ref{eq:utility_maximizer_nonlinear}.
Suppose we observe the radar over $n=1,2,\ldots,N$ time epochs (slow varying time scale). At the $n^{th}$ epoch, we probe the radar with an interference vector $\alpha_n\in\reals^M$.
The radar responds with waveform $\beta_n\in\reals_{+}^M$. We assume the chosen waveform $\beta_n$ maximizes the radar's underlying utility function while ensuring the radar's SINR exceeds a particular threshold $\delta>0$, where the SINR of the radar given probe $\probe$ and response $\response$ is defined as
\beq \label{eqn:SINR_def}
\operatorname{SINR}(\alpha,\beta) = \frac{\beta^{'}Q\beta}{\beta^{'}P(\alpha)\beta + \gamma}.
\eeq
In~(\ref{eqn:SINR_def}), the radar's signal power (numerator) and interference power (first term in denominator) are assumed to be quadratic forms of $Q,P(\probe)$ respectively, where $Q,P(\alpha)\in\reals^{M\times M}$ are positive definite matrices known to us. The term $\gamma>0$ is the noise power. The SINR definition in~(\ref{eqn:SINR_def}) is a more general formulation of the $\operatorname{SCNR}$~(\ref{Eq:SINR}) of a cognitive radar  derived in Sec.\,\ref{sec:Interference} using clutter response models~\cite{Guerci16}. The matrices $Q,P(\alpha)$ are analogous to the covariance of the channel impulse response matrices $H_t(\cdot)$ and $H_p(\cdot)$ corresponding to the target and clutter~(external interference) channels, respectively~(see Sec.\,\ref{sec:signal_interference} for a  discussion).

Having defined the SINR above in \eqref{eqn:SINR_def}, we now formalize the radar's response $\response_n$ given probe $\probe_n$, $n=1,2,\ldots$ as the solution of the following constrained optimization problem.
\begin{align}
    \beta_n &\in\argmax_{\beta} U(\beta)\nonumber\\
    \text{s.t. }&\operatorname{SINR}(\alpha_n,\beta) \geq \delta\label{eqn:radar_opt_nonlinear}
\end{align}
Clearly, the above setup falls under the non-linear utility maximization setup in Definition~\ref{eq:utility_maximizer_nonlinear}
by defining the non-linear budget $g_n(\cdot)$  as $g_n(\beta)=\delta - SIR(\alpha_n,\beta)$ where $SIR(\cdot)$ is defined in~(\ref{eqn:SINR_def}). It only remains to show that this definition of $g_n(\beta)$ is monotonically increasing in $\beta$. Theorem~\ref{thm:sufficient} stated below establishes two conditions that are sufficient for $g_n(\beta)$ to be monotonically increasing in $\beta$.
\begin{theorem}
Suppose that the adversary radar uses the SINR constraint \eqref{eqn:radar_opt_nonlinear}. Then $g_n(\beta)=\delta-\operatorname{SIR}(\alpha_n,\beta)$ is monotonically increasing in $\beta$ if the following two conditions hold.
\begin{compactenum}
    \item The matrix $Q$ is a diagonal matrix with off-diagonal elements equal to zero. 
    \item The matrix $\left(\frac{c_{P(\alpha_n)}}{d_Q}P(\alpha_n) - Q\right)$ is component-wise less than $0$ for all $n\in\{1,2,\ldots N\}$, where $c_{P(\alpha_n)}>0$ and $d_Q>0$ denote the smallest and largest eigenvalues of $P(\alpha_n)$ and $Q$ respectively.
\end{compactenum}
\label{thm:sufficient}
\end{theorem}
The proof of Theorem \ref{thm:sufficient} follows from elementary calculus and omitted for brevity. Hence, assuming the two conditions hold in Theorem~\ref{thm:sufficient} above, we can use the results from Theorem~\ref{thm:nonlinearAfriat} to test if the radar satisfies utility maximization in its waveform design~(\ref{eqn:radar_opt_nonlinear}) and also estimate the set of feasible utility functions $U(\cdot)$~(\ref{eqn:radar_opt_nonlinear}) that rationalizes the radar's responses $\{\beta_n\}$.

\section{Designing  Smart Interference To Confuse Cognitive Radar}\label{sec:Interference}
 \begin{figure*}
    \centering
            {\resizebox{15cm}{!}{
\begin{tikzpicture}[node distance = 1cm, auto]
    \node [blocka] (BLOCK1) {Receiver};
    \node [blocka, below of=BLOCK1,right of=BLOCK1,node distance=1.5cm] (BLOCK2) {Transmitter};
    \node [blocka, below of=BLOCK1,left of=BLOCK1,node distance=1.5cm] (BLOCK3) {Tracker\\(Estimator)};

    \draw[<-] (BLOCK1) -| node[left,pos=0.8]{Waveform $\vect W$}  (BLOCK2)  ;
    \draw[->] (BLOCK1.west) -|   node[left,pos=0.8]{$\vect X$} (BLOCK3);

    \draw[->](BLOCK3) --  node[above]{\null} (BLOCK2);
    
    \node [blocka,right of=BLOCK1,node distance = 8.5cm] (BLOCK4) {Transmitter};
    \node [blocka, below of=BLOCK4,right of=BLOCK4,node distance=1.5cm] (BLOCK5) {Decision Maker};
    \node [blocka, below of=BLOCK4,left of=BLOCK4,node distance=1.5cm] (BLOCK6) {Receiver};

    \draw[<-] (BLOCK4) -| node[left,pos=0.8]{$\vect H_p$}  (BLOCK5)  ;
    \draw[->] (BLOCK4.west) -|   node[left,pos=0.8]{\null} (BLOCK6);

    \draw[->](BLOCK6) --  node[above]{$\vect Y$} (BLOCK5);
    \node [blocka, above of=BLOCK1,node distance = 2cm] (DRONE) {Tracked Target};
    \draw[<->,dashed,line width=2pt] ([xshift=-0.6cm]BLOCK1.north) -- node[left]{Transmit channel $\vect H_t$} ([xshift=-0.6cm]DRONE.south);
    \draw[<->,dashed,line width=2pt] ([xshift=0.6cm]DRONE.south) -- node[right]{Clutter channel $\vect H_c$} ([xshift=0.6cm]BLOCK1.north);

    \draw[->,color=red,line width=2pt] ([xshift=0.2cm]BLOCK2.east)   -- node[above]{Noisy measurement $\vect Y$}node[below]{of transmit waveform $\vect W$}([xshift=-0.2cm]BLOCK6.west);
    \draw[->,line width=2pt] ([xshift=-1.2cm]BLOCK4.west)   --   node[below]{Interference Signal $\vect H_p$} ([xshift=1.2cm]BLOCK1.east);
    \node[draw] at (8.5,-2.5) {{\color{blue}Our side}};
    \node[draw] at (0,-2.5) {{\color{red}Adversary}};
   \end{tikzpicture}} }

\caption{Schematic of transmit channel $H_t$, clutter channel $H_c$ and interference signal $\vect H_p$ involving an adversarial cognitive radar and us. We observe the radar's waveform $\vect W$ in noise. The aim is to engineer the interference signal $\vect H_p$ to confuse the adversary cognitive radar.}
\label{Fig:CoFAR}
\end{figure*}
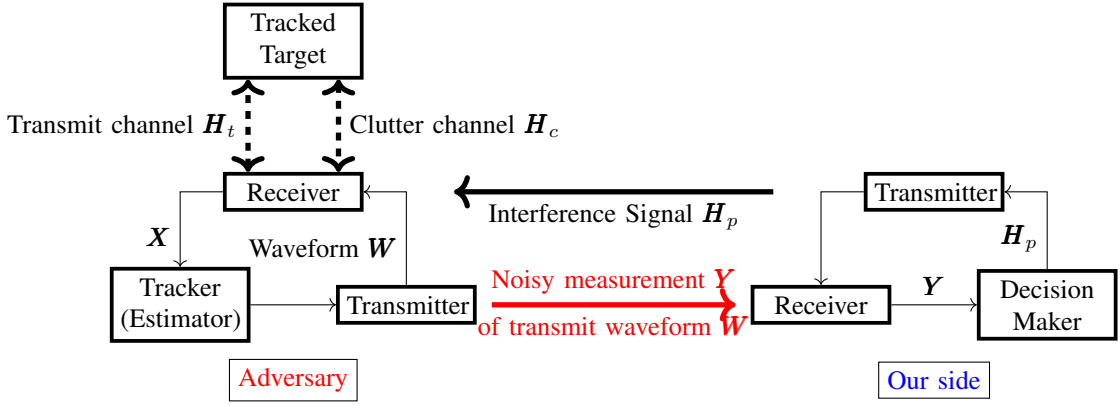

This section discusses how we can engineer (design)  external interference (a probing signal) at the physical layer level to confuse a cognitive radar. By abstracting the probing signal to a channel in the frequency domain, our objective is to minimize the signal power of the interference generated by us while ensuring the $\operatorname{SCNR}$ of the radar does not exceed a pre-defined threshold. The setup is schematically shown in Figure~\ref{Fig:CoFAR}. Note that the level of abstraction used in this section is at the Wiener filter pulse/waveform level; whereas the previous two sections were at the systems level (which uses the utility maximization framework) and tracker level (which uses the Kalman filter formalism), respectively. This is consistent with the design theme of sense globally (high level of abstraction)  and act locally (lower level of abstraction).

As can be seen in the $\operatorname{SCNR}$ expression~(\ref{Eq:SINR}), the interference signal power manifests as additional clutter perceived by the radar in the denominator thus forcing the $\operatorname{SCNR}$ to go down. The radar then re-designs its waveform to maximize its $\operatorname{SCNR}$ given our interference signal. We observe the adversarial radar's chosen waveform in noise. Our task can thus be re-formulated as choosing the interference signal with minimal power while ensuring that with probability at least $1-\epsilon$, the optimized $\operatorname{SCNR}$ lies below a threshold level $\Delta$. Here , $\epsilon$ and $\Delta$ are user-defined parameters.  
This approach closely follows the formulation in Sec.\,\ref{sec:waveform} where the cognitive radar chooses the optimal waveform while ensuring the $\operatorname{SINR}$ exceeds a threshold value. Further, the SCNR of the adversary's radar defined in \eqref{eqn:SCNR_def} below can be interpreted as a monotone function of the radar's utility function in the abstracted setup of Sec.\,\ref{sec:waveform}, since in complete analogy to the utility maximization model of Sec.\,\ref{sec:waveform}, this section assumes the radar maximizes its SCNR in the presence of smart interference signals (probes).

\subsection{Interference Signal Model}\label{sec:signal_interference}
We first characterize how a cognitive radar optimally chooses its waveform based on its perceived interference. The radar's objective is to choose the optimal waveform that maximizes its signal-to-interference-plus-noise~($\operatorname{SINR}$) ratio. 


Suppose we observe the adversarial radar over $l=1,2\ldots L$ pulses, where each pulse comprises $n=1,2,\ldots N$ discrete time steps.
A single-input single-output (SISO) radar system has two channel impulse responses, one for the target and the other for clutter. Let $w(n)$ denote the  radar transmit waveform and $h_t(n)$, $h_c(n)$ denote the target and clutter channel impulse responses, respectively. Then, the radar measurements corresponding to the $l$-th pulse can be expressed as
\begin{equation}
x(n,l) = h_t(n,l) \circledast w(n,l) + h_c(n,l) \circledast w(n,l) + e_r(n,l)
\end{equation}
where $\circledast$ represents a convolution operator and $e_r(n,l)$ is the radar measurement noise modeled as an i.i.d random variable with zero mean and known variance $\sigma_r^2$. We model the radar's measurement using the stochastic Green's function impulse response model presented in \cite{Guerci16}, where the radar's electromagnetic channel is modeled using a physics based impulse response. 

Since convolution in the time domain can be expressed as multiplication in the frequency domain (with notation in upper case), we can express the measurements in the frequency domain  as follows:
\begin{equation}
X(k,l) = H_t(k,l) W(k,l) + H_c(k,l) W(k,l) + E_r(k,l)
\label{Eq:Freq_domain}
\end{equation}
where $k\in\mathcal{K}=\{1,\ldots,K\}$ is the frequency bin index.
Eq. \eqref{Eq:Freq_domain} can be extended to an $I \times J$ MIMO radar and the received signal at the $j$-th receiver is given by
\begin{align}
  X_j(k,l) = \sum_{i=1}^I& H_{t_{ij}}(k,l) W_i(k,l) +H_{c_{ij}}(k,l) W_i(k,l) + E_{r,j}(k,l),\label{Eq:Freq_domain_MIMO}
\end{align}
$\forall k\in\{1,\ldots K\}$. Using matrices and vectors obtained by stacking and concatenating  \eqref{Eq:Freq_domain_MIMO} for all $i,j,$ and $k$, the MIMO radar measurement model at the $l^{th}$ pulse in vector-matrix form can be expressed as
\begin{equation}\label{eqn:recvec_interference}
\vect X(l) = \vect H_t(l)\vect W(l) + \vect H_c(l) \vect W(l) + \vect E_r(l)
\end{equation}
where $\vect X(l) \in \mathbb{C}^{(J\times K)\times 1}$ is the received signal vector, $\vect H_c(l),\vect H_t(l)\in\mathbb{C}^{(J\times K)\times (I\times J\times K)}$ are the effective transmit and clutter channel impulse response matrices respectively, $\vect W(l)\in\mathbb{C}^{(I\times J\times K)\times 1}$ is the radar's effective waveform vector. $\vect E_r(l) \in \mathbb{C}^{(J\times K)\times 1}$ is the effective additive noise vector modeled as a zero mean i.i.d random variable (independent over pulses) with covariance matrix $C_r\in\mathbb{R}^{(J\times K)\times(J\times K)},C=(\sigma_r^2/K)\vect I=\tilde{\sigma}_r^2\vect I$. The block diagram in Fig. \ref{Fig:CoFAR} shows the entire procedure for this model.

\subsection{Smart Interference to confuse the adversary radar}
The aim of this section is to design optimal interference signals (to confuse the adversary cognitive radar) by solving a probabilistically constrained optimization problem.

 At the beginning of the $l^{th}$ pulse, the adversary  radar transmits a pilot signal to estimate the transmit and clutter channel impulse responses $\vect H_t(l)$ and $\vect H_c(l)$ respectively. Assuming it has a perfect estimate of $\vect H_t(l)$ and $\vect H_c(l)$, the radar then chooses the optimal waveform $\vect W^{\ast}(l)$ such that $\operatorname{SCNR}$ defined below in~(\ref{Eq:SINR}) is maximized. The radar's waveform $\vect W^{\ast}(l)$ is the solution to the following optimization problem
\begin{align}
\vect W^{\ast}(l) &= \argmax_{\vect W (l): \norm{\vect W(l)}_2 = 1} \operatorname{SCNR}(\vect H_t(l),\vect H_c(l),\vect W(l)),
\label{Eq:SINR}
\end{align}
where the SCNR is defined as
\beq \label{eqn:SCNR_def}
\operatorname{SCNR}(\vect H_t, \vect H_c, \vect W ) = \dfrac{ \big\Arrowvert \vect H_t \vect W\big\Arrowvert_2^2 }{\mathbb{E} \Big\{ \big\Arrowvert\vect H_c \vect W + \vect E_r\big\Arrowvert_2^2\Big\}}.
\eeq
 Denote the maximum $\operatorname{SCNR}$ achieved in~(\ref{Eq:SINR}) as 
\begin{equation}\label{eqn:max_SCNR}
    \operatorname{SCNR}_{\max}(\vect H_t(l),\vect H_c(l),\sigma_r^2) = \operatorname{SCNR}(\vect H_t(l),\vect H_c(l),\vect W^{\ast}(l)).
\end{equation}
Given $\vect H_t(l),\vect H_c(l)$ and the radar's measurement noise power $\sigma_r^2$, the radar generates an optimal waveform at the $l^{th}$ pulse using \eqref{Eq:SINR} as
 the solution to the following eigenvector problem~\cite{BGG15}:
\begin{align}
\vect A~\vect W^\star (l) &= \lambda_l \vect W^\star (l)\nonumber\\
\vect A&=\Big(\big(\vect H_c(l)' \vect H_c(l) + \tilde{\sigma}_r^2 \vect I\big)^{-1} \vect H_t(l)' \vect H_t(l)\Big),\nonumber
\end{align}
Here $(\cdot)'$ denotes the Hermitian transpose operator.

\begin{figure}[h]
	\centering
	\includegraphics[scale=0.34]{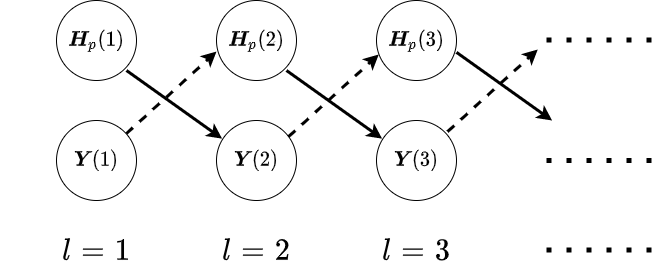}
	\caption{Schematic of smart interference design to confuse the cognitive radar. The interference signal at the $l^{th}$ pulse affects the waveform choice of the radar in the $(l+1)^{th}$ pulse. We record the noisy waveform measurement $\vect Y(l+1)$ and generate the interference signal for the $(l+2)^{th}$ pulse.}\label{fig:dyn_interference}
\end{figure}

As an external observer, we send a sequence of probe signals $P=\{\vect H_p(l),l\in\{1,2,\dots L\}\}$ over $L$ pulses to confuse the adversary radar and degrade its performance. The interference signal $\vect H_p(l-1)$ at the $(l-1)^{th}$ affects only radar's clutter channel impulse response $\vect H_c(l)$ at the $l^{th}$ pulse which subsequently results in change of optimal waveform~(\ref{Eq:SINR}) chosen by the radar $\vect W^{\ast}(l)$. We measure the optimal waveform at the $l^{th}$ pulse in noise as $\vect Y(l)$. We assume constant transmit and clutter channel impulse responses $\vect H_t,\vect H_c$ in the absence of the probe signals $P$. The dynamics of our interaction with the adversary radar due to probe $P$ are as follows and schematically shown in Fig.\,\ref{fig:dyn_interference}:
\begin{align}
\vect H_c(l)  &=  \vect H_c + \vect H_p(l-1)\label{Eq:H_c}\\
\vect H_t(l) &= \vect H_t \\ 
\Big(\big(\vect H_c(l)' &\vect H_c(l) + \tilde{\sigma}_r^2 \vect I\big)^{-1} \vect H_t(l)' \vect H_t(l)\Big)\vect W^{\ast} (l) \nonumber\\ 
 &= \lambda_l \vect W ^{\ast}(l)\label{Eq:Eigen}\\
\vect Y(l) &=\vect W^*(l) + \vect E_o(l).
\label{Eq:Y}
\end{align}
In~(\ref{Eq:Y}), $E_o(l)$ is our measurement noise modeled as a zero mean i.i.d random variable~(independent over pulses) sampled from a known pdf $f_o$ with zero mean and covariance $C_o=(\sigma_o^2/K)\vect I = \tilde{\sigma}_o^2\vect I$. 

Our objective is to optimally design the probe signals $P^{\ast}=\{\vect H_p^{\ast}(l),l\in\{1,\ldots L\}\}$ that minimizes the interference signal power such that for a pre-defined $\Delta>0$, there exists $\epsilon\in[0,1)$ such that the probability the $\operatorname{SCNR}$ of the radar lies below $\Delta$ exceeds $(1-\epsilon)$, for all $l=1,2,\ldots L$.
\begin{align}
\min_{\{\vect H_p(l),l\in\{1,2,\ldots L\}\}} & \sum_{l=1}^L \vect H_p(l)'\vect H_p(l)\nonumber\\
 \text{s.t. } \mathbb{P}_{f_o}(\operatorname{SCNR}&(\vect H_t(l),\vect H_c(l),\vect Y(l)) \leq \Delta )\geq 1-\epsilon,\nonumber\\
 &\forall l\in\{1,2,3,\ldots L\}.\label{eqn:opt_us_interference}
\end{align}
Here, $\mathbb{P}_{f}(\cdot)$ denotes the probability  wrt pdf $f$. The design parameter $\Delta$ is the SCNR (performance) upper bound of the cognitive radar. To confuse the radar, our task is to ensure the SCNR of the radar is less than $\Delta$ with probability at least $1-\epsilon$. Hence, $\epsilon$ is the maximum probability of failure to confuse the radar with  our smart interference signals.
Although not explicitly shown, the $\operatorname{SCNR}_{\max}$ expression in~(\ref{eqn:opt_us_interference}) depends on our interference signal $\vect H_p$ as depicted  in~(\ref{Eq:H_c}).

Solving the non-convex optimization problem \eqref{eqn:opt_us_interference} is challenging except for trivial cases. It involves two inter-related components $(i)$ Estimating the transmit and clutter channel impulse responses $\vect H_t,\vect H_c$ from observation $\vect Y(l)$ and $(ii)$ Using the estimated value of channel impulse responses to generate the interference signal $\vect H_p(l)$. 
Moreover, solving for $\vect H_c$ and $\vect H_t$ from recursive equations \eqref{Eq:H_c} through \eqref{Eq:Y} for $l=1,\ldots,L$ is a challenging problem since it does not have an analytical closed form solution.

With the above formulation, we can now discuss construction of smart inference to confuse the adversary  radar.
The cognitive radar maximizes its energy in the direction of its target impulse response and transfer function. As soon as we have an accurate estimate of the target channel transfer function from the $L$ pulses, we can immediately generate signal dependent interference that nulls the target returns. Even if the clutter channel impulse response changes after we perform our estimation, since the target channel is stationary for longer durations, the signal dependent interference will work successfully for several pulses after we compute the estimate. The 
main take away from this approach is that we are exploiting the fact that the cognitive radar provides information about its channel by optimizing the waveform with respect to its environment. 

\subsection{Numerical example illustrating design of smart interference}
We conclude this section with a numerical example that illustrates the smart interference framework developed above. The simulation setup is as follows:
\begin{compactitem}
\item $L=2$ pulses (optimization horizon in \eqref{eqn:opt_us_interference}).
\item Impulse response matrices for transmit channel $\vect H_t=[7~7]$, clutter channel $\vect H_c=[1~1]$, and adversary radar noise covariance $\tilde{\sigma}_r^2=1$ \eqref{eqn:recvec_interference}. 
\item {\em Design parameters:} SCNR upper bound $\Delta=\{2.8,3,3.2\}$, minimum probability of success\\ $\epsilon=0.2,0.3$ \eqref{eqn:opt_us_interference}.
\item Probe signals for pulse index: \\ 
\beq \label{eqn:parametrization}\begin{split}
&l=1,~\vect H_p(1)=[0.2r~0.5r], \\
&l=2,~\vect H_p(2)=[0.4r~0.4r].\end{split} 
\eeq  The smart interference parameter $r>0$ parametrizes the magnitude of the probe signals . The aim is to find the optimal   probe signals $\vect H_p(l),~l=1,2$ parametrized by $r$ in \eqref{eqn:parametrization} that solves \eqref{eqn:opt_us_interference}. The corresponding value of $r$ is our optimal smart interference parameter.
\item Our measurement noise covariance is $\tilde{\sigma}_o^2=0.1$ \eqref{Eq:Y}.
\end{compactitem}

\begin{figure}[h]
    \centering
    \includegraphics[width=\linewidth]{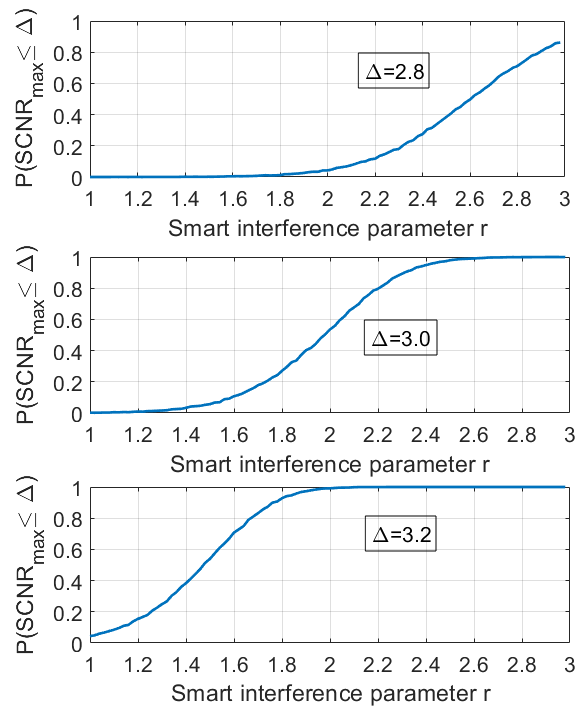}
    \caption{
    The figure illustrates the performance of the cognitive radar as our smart interference parameter $r$  in  \eqref{eqn:parametrization} is varied.
    The plots  display  the LHS of \eqref{eqn:opt_us_interference}, namely,  probability that the radar's maximum SCNR (which depends on $r$ \eqref{eqn:parametrization}) is smaller than threshold  $\Delta$. The probability curves are plotted for $\Delta=2.8,3,3.2$ and signify the extent of SCNR degradation as a function of the magnitude of the probe signal.} 
    \label{fig:interference_sim}
\end{figure}

Figure \ref{fig:interference_sim} displays the performance of the cognitive radar as our smart interference parameter $r$ is varied. It shows that increasing $r$ leads to increased confusion (worse $\operatorname{SCNR}$ performance) of the cognitive radar.
Specifically, we plot the LHS of \eqref{eqn:opt_us_interference}, namely, $\mathbb{P}_{f_o}(\operatorname{SCNR}(\vect H_t(l),\vect H_c(l),\vect Y(l)) \leq \Delta)$, for SCNR upper bound $\Delta\in\{2.8,3,3.2\}$.  
Recall that this is the probability with which the maximum SCNR of the radar \eqref{eqn:max_SCNR} lies below~$\Delta$.

To glean insight from Figure \ref{fig:interference_sim}, let $r^{\ast}(\Delta, \epsilon)$ denote the optimal smart interference parameter  that  solves \eqref{eqn:opt_us_interference} for design parameters $\Delta$ and $\epsilon$. Figure \ref{fig:interference_sim} shows that $r^{\ast}(\Delta, \epsilon)$ decreases with both design parameters $\Delta$ and $\epsilon$.
This can be justified as follows. For a fixed value of failure probability $\epsilon$, increasing the upper bound $\Delta$ implies
the constraint \eqref{eqn:opt_us_interference} is satisfied for smaller $r$, hence the optimal interference parameter $r^{\ast}(\Delta, \epsilon)$ decreases with $\Delta$. Recall  $\epsilon$ upper bounds the probability with which the maximum SCNR of the radar exceeds $\Delta$. Increasing $\epsilon$ (or equivalently, relaxing the maximum probability of failure) allows us to decrease the magnitude of the probe signals without violating the constraint in \eqref{eqn:opt_us_interference} for a fixed $\Delta$. Hence, $r^{\ast}(\Delta, \epsilon)$ decreases with both $\Delta$ and $\epsilon$.



\section{Conclusion}
This paper considered three important inter-related aspects of adversarial inference involving cognitive radars. First we discussed inverse tracking (estimating the adversary  tracker's estimate based on the radar's actions) and calibration of the adversary's sensor accuracy. Then we presented a revealed preferences methodology for identifying cognitive radars; i.e.\,, identifying a constrained utility maximizer. Finally, we discussed designing interference to confuse the cognitive radar. The above three aspects are inter-related as depicted in Figure \ref{fig:blockdiag}. The levels of abstraction range from smart interference design based on Wiener filters (at the pulse/waveform level), inverse Kalman filters at the tracking level and revealed preferences for identifying utility maximization at the systems level.

\subsubsection*{Extensions}
The results in this paper lead to several interesting future extensions. There is strong motivation to determine analytic performance bounds for inverse tracking/filtering and estimation of the adversary's sensor gain. Another aspect (not considered here) is when the adversary does not know the transition kernel of our  dynamics; the adversary then needs to estimate this transition kernel, and we need to estimate the estimate of this transition kernel.
In future work we will design the smart interference problem \eqref{eqn:opt_us_interference} as a  stochastic control problem; since dynamic programming is intractable we will explore limited look-ahead policies and open-loop feedback control.

Regarding identifying cognitive radars, it is worthwhile  developing statistical tests for utility maximization when the response of the utility maximizing adversarial radar is observed in noise; see Varian's work \cite{Var06} on noisy revealed preference. Ongoing research in developing a dynamic revealed preference framework will be used to extend the beam allocation problem of Sec.\,\ref{sec:beam} to a multi-horizon setup where we analyze batches of adversary responses over multiple slow time scale epochs.
Another natural extension is to a Bayesian context, namely, identifying a radar that is a Bayesian utility maximizer. We refer to \cite{CD15} for seminal work in this area stemming from behavioral economics.

Finally, in the design of controlled inference, it is worthwhile considering a game-theoretic setting where the cognitive radar (adversary) and us interact dynamically. In previous work \cite{MK07b} we studied simpler versions of the problem in the context of cross-polarized jamming. Also, in future work, it is worthwhile to develop a stochastic gradient algorithm for estimating the optimal probe signal.


\bibliographystyle{abbrv}

\bibliography{vkm}
\begin{IEEEbiography}[{\includegraphics[width=1in,height=1.25in,clip,keepaspectratio]{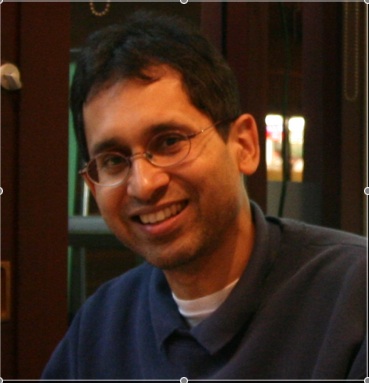}}]{Vikram Krishnamurthy}
(F'05) received the Ph.D. degree from the Australian National University
in 1992. He is  a professor in the School of Electrical \& Computer Engineering,
Cornell University. From 2002-2016 he was a Professor and Canada Research Chair
at the University of British Columbia, Canada. His research interests include statistical signal processing  and stochastic control in social networks and adaptive sensing. He served as Distinguished Lecturer for the IEEE Signal Processing Society and Editor-in-Chief of the IEEE Journal on Selected Topics in Signal Processing.
In 2013, he was awarded an Honorary Doctorate from KTH (Royal Institute of Technology), Sweden. He is author of two books {\em Partially Observed Markov Decision Processes} and {\em Dynamics of Engineered Artificial Membranes and Biosensors} published by Cambridge University Press in 2016 and 2018, respectively.
\end{IEEEbiography}

\begin{IEEEbiography}[{\includegraphics[width=1in,height=1.25in,clip,keepaspectratio]{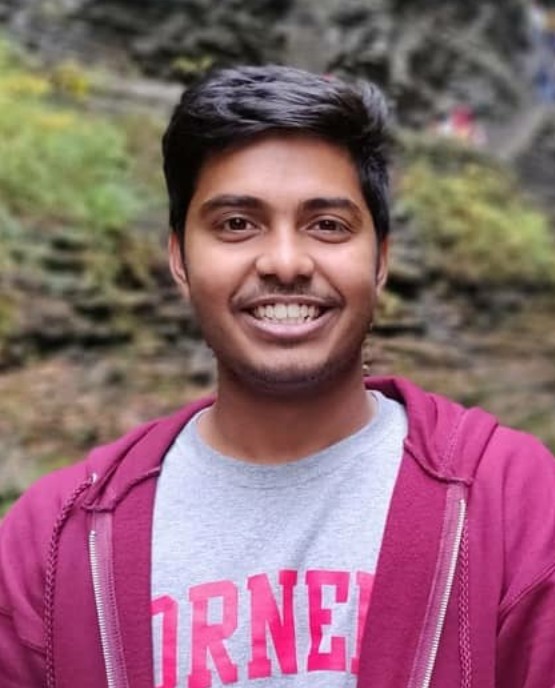}}]{Kunal Pattanayak} (S'21) received the integrated Bachelors and Masters in Technology degrees in Electronics and Electrical Communication  Engineering from Indian Institute of Technology, Kharagpur in 2018. He is currently a graduate student in the department of Electrical and Computer Engineering at Cornell University. His research interests include inverse reinforcement learning, behavioral economics, statistical signal processing, design of counter-autonomous systems for radar applications and interpretable AI. He is a recipient of the McMullen graduate fellowship by Cornell University, and has been a speaker at the 2020 Sloan-NOMIS Conference on Attention and Applied Economics.  
\end{IEEEbiography}

\begin{IEEEbiography}[{\includegraphics[width=1in,height=1.25in,clip,keepaspectratio]{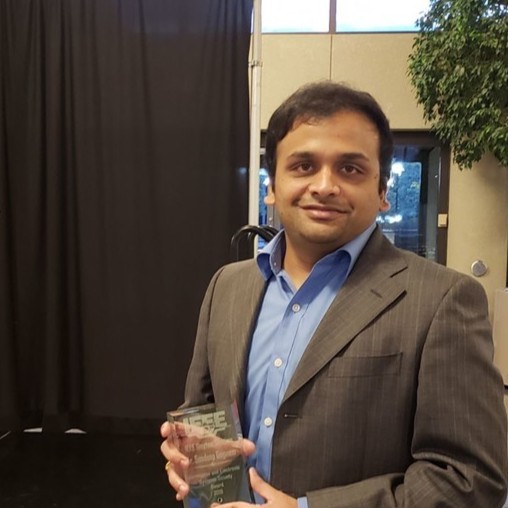}}]{Sandeep Gogineni}
(BTech, Electronics and Communications Engineering, IIIT, India, MS, PhD, Electrical Engineering, Washington University in St. Louis) is a Research Scientist for Information Systems Laboratories with over 12 years of experience working on radar and wireless communications systems. He has worked for 6 years as an on-site contractor for Air Force Research Laboratory (AFRL), developing novel signal processing algorithms and performance analysis for passive radar systems. He received the IEEE Dayton Section Aerospace and Electronics Systems Society Award for these contributions to passive radar signal processing. Prior to his time at AFRL, during his graduate studies at Washington University in St. Louis, Dr. Gogineni developed optimal waveform design techniques for adaptive MIMO radar systems and demonstrated improved target detection and estimation performance. At ISL, Dr. Gogineni has been working on channel estimation algorithms and optimal probing strategies for MIMO radar systems in the context of Cognitive Fully Adaptive Radar (CoFAR). Additionally, Dr. Gogineni and his colleagues at ISL have demonstrated the feasibility of using neural networks and artificial intelligence techniques to solve extremely challenging radio frequency (RF) sensing problems. His expertise includes statistical signal processing, detection and estimation theory, deep learning, artificial intelligence, performance analysis, and optimization techniques with applications to active and passive RF sensing systems.
\end{IEEEbiography}

\begin{IEEEbiography}[{\includegraphics[width=1in,height=1.25in,clip,keepaspectratio]{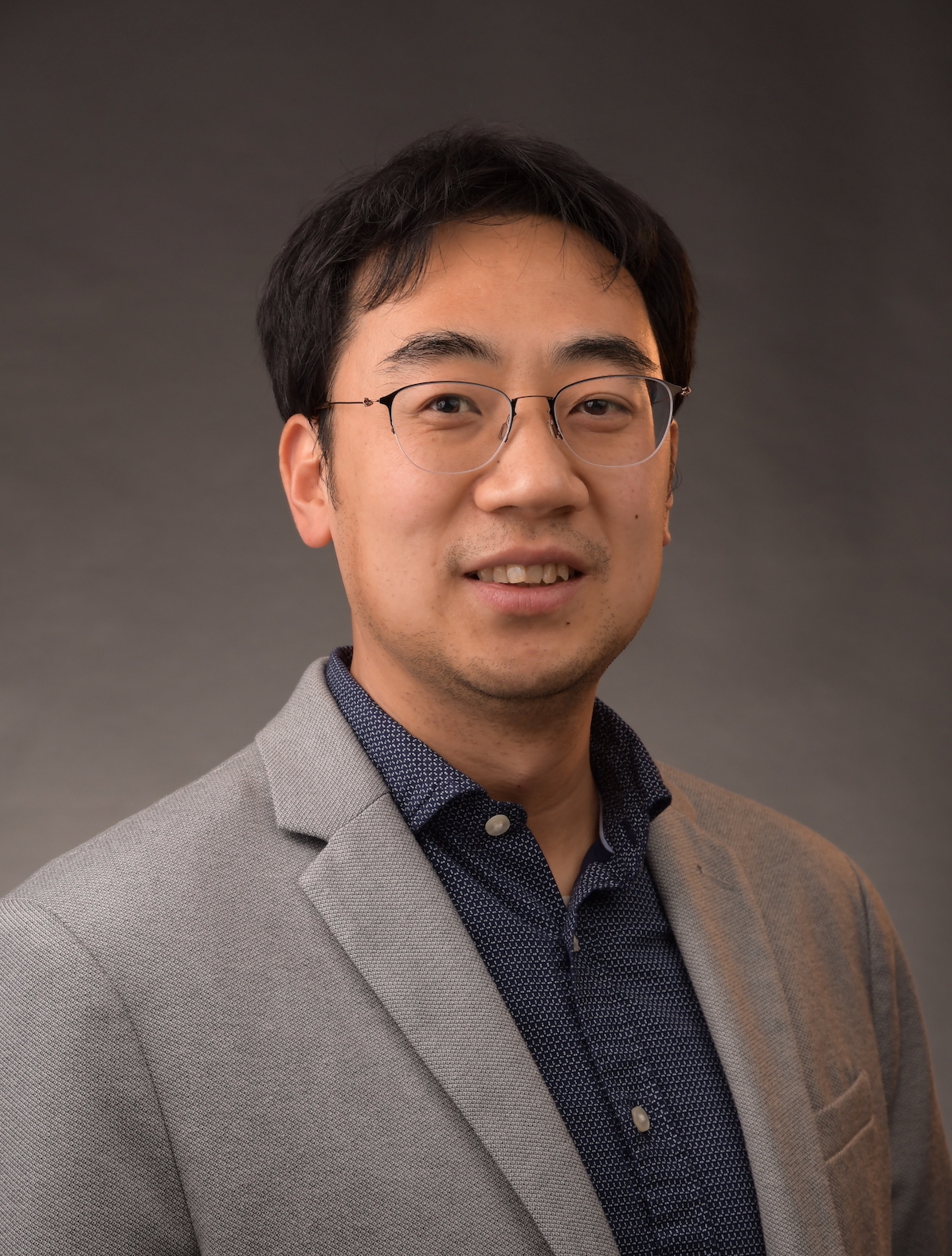}}]{Bosung Kang} (S'12-M'16) received the B.S. and M.S. degrees in Electrical and Electronic
Engineering from Yonsei University, Seoul, South Korea, in 2005 and 2007, respectively, and the
Ph.D. degree in Electrical Engineering from the Pennsylvania State University, University Park,
PA, USA, in 2015. He is currently an onsite contractor for Air Force Research Laboratory,
WPAFB, OH and a radar research engineer at University of Dayton Research Institute, Dayton,
OH.
He worked at LG Electronics as a research engineer, Seoul, Korea, from 2007 to 2011. He
developed image and video signal processing algorithms in mobile camera and monitor
applications. He has served as a reviewer for several reputed IEEE journals and conferences. His
research interests include statistical signal processing, detection and estimation, convex
optimization, and radar signal processing with applications to radar and communication
systems.
Dr. Kang was a recipient of the First Place in the Student Paper Competition at the IEEE Radar
Conference, Cincinnati, OH, in 2014. He also won the 2015 Robert T. Hill Best Dissertation
Award presented by the IEEE Aerospace and Electronic Systems Society (AESS).
\end{IEEEbiography}

\begin{IEEEbiography}[{\includegraphics[width=1in,height=1.25in,clip,keepaspectratio]{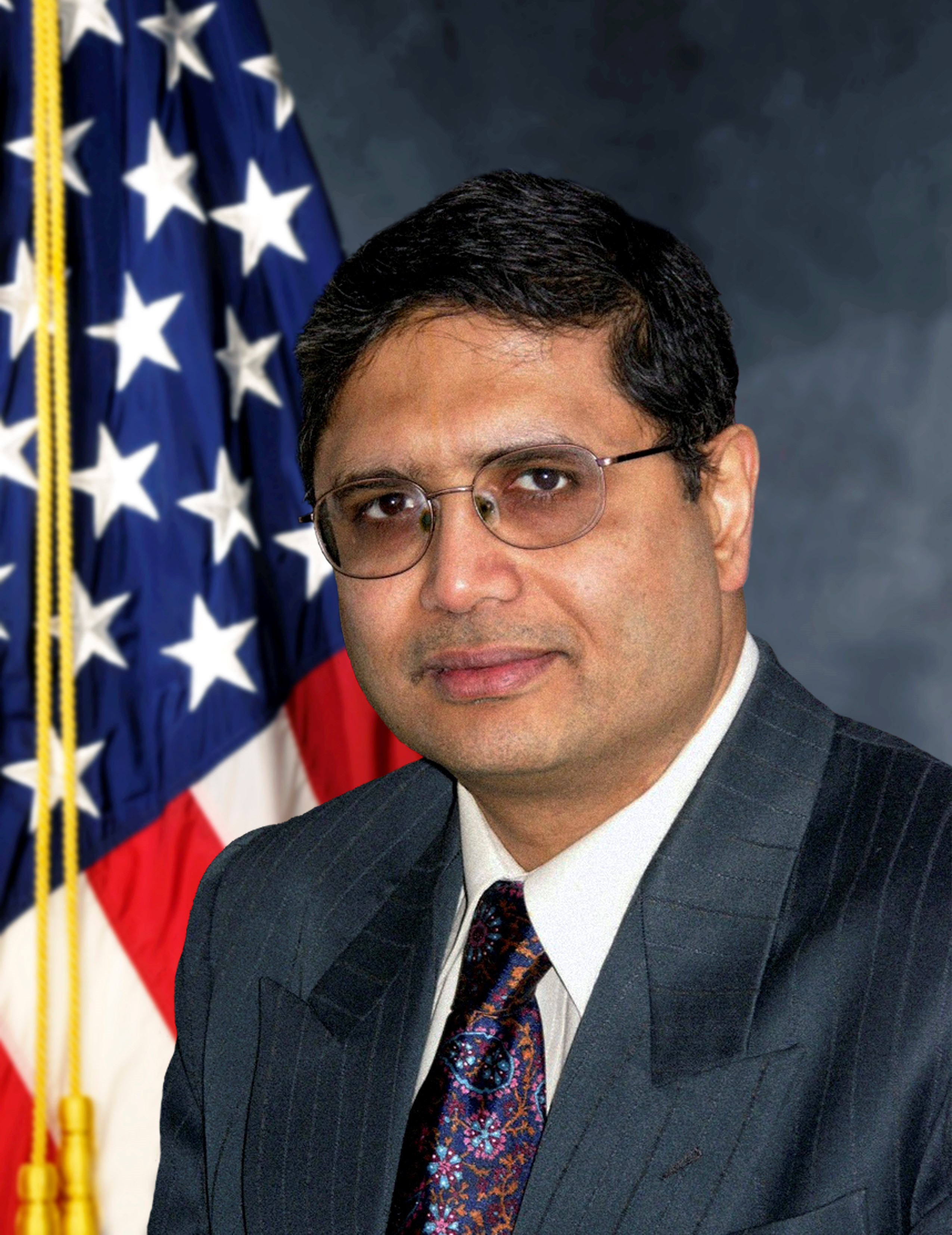}}]{Muralidhar Rangaswamy}
(Fellow, IEEE) received the B.E. degree in electronics engineering from Bangalore University, Bangalore, India, in 1985, and the M.S. and
Ph.D. degrees in electrical engineering from Syracuse University, Syracuse, NY, USA,
concurrently in 1992.
He is currently the Technical Lead for Radar Sensing with the RF Technology Branch
within the Sensors Directorate, Air Force Research Laboratory (AFRL), Wright-Patterson
Air Force Base, OH, USA. Prior to this, he has held industrial and academic appointments.
He is a contributor to eight books and is a co-inventor on three U.S. patents. He has
coauthored more than 280 refereed journal and conference record papers in the areas of his
research interests, which include radar signal processing, spectrum estimation, modeling
non-Gaussian interference phenomena, and statistical communication theory.
Dr. Rangaswamy is a member of the Radar Systems Panel (RSP) in the IEEE-AES
Society. He was on the Technical Committee of the IEEE Radar Conference series in a
myriad of roles. He was the recipient of the IEEE Warren White Radar Award in 2013,
the 2013 Affiliate Societies Council Dayton (ASC-D) Outstanding Scientist and Engineer
Award, the 2007 IEEE Region 1 Award, the 2006 IEEE Boston Section Distinguished
Member Award, and the 2005 IEEE-AESS Fred Nathanson Memorial Outstanding Young
Radar Engineer Award. He was also the recipient of the 2012 and 2005 Charles Ryan Basic
Research Award from the Sensors Directorate, AFRL, in addition to more than 40 scientific
achievement awards. Most recently, he received the International Society for Information
Fusion Jean-Pierre Le Cadre Best Paper Award at the 2019 FUSION Conference, the 2019
Technical Cooperation Panel Award from the Office of Secretary of Defense, and the 2019
IEEE Dayton Section Fritz Russ Memorial Award.
\end{IEEEbiography}

\end{document}